\documentclass[
    a4paper,
    prb,
    bibnotes,
    twocolumn,
    superscriptaddress,
    groupedaddress,
    showpacs,
]{revtex4}
\usepackage{graphicx}
\usepackage{amssymb,amsmath}
\usepackage{amsmath,mathrsfs}

\newcommand{\Eqref}[1]{Eq.~(\ref{#1})}
\newcommand{\Figref}[1]{Fig.~\ref{#1}}
\newcommand{\Secref}[1]{Sec.~\ref{#1}}

\begin{document}

\title{Current-induced atomic dynamics, instabilities, and Raman signals:\\
Quasi-classical Langevin equation approach}
 \author{Jing-Tao
  \surname{L\"u}} \email{jtlu@nanotech.dtu.dk} \affiliation{DTU
  Nanotech, Department of Micro- and Nanotechnology, Technical
  University of Denmark, {\O}rsteds Plads, Build. 345E, DK-2800
  Kongens Lyngby, Denmark}
\author{Mads \surname{Brandbyge}}
\email{Mads.Brandbyge@nanotech.dtu.dk} \affiliation{DTU Nanotech,
  Department of Micro- and Nanotechnology, Technical University of
  Denmark, {\O}rsteds Plads, Build. 345E, DK-2800 Kongens Lyngby,
  Denmark}
\author{Per \surname{Hedeg{\aa}rd}} \email{hedegard@fys.ku.dk}
\affiliation{Niels Bohr Institute, Nano-Science Center, University of
  Copenhagen, Universitetsparken 5, 2100 Copenhagen {\O}, Denmark}
\author{Tchavdar N. Todorov} \author{Daniel Dundas}
\affiliation{Atomistic Simulation Centre, School of Mathematics and
  Physics, Queen's University Belfast, Belfast BT7 1NN, United
  Kingdom} \date{\today}

\begin{abstract}
  We derive and employ a semi-classical Langevin equation obtained
  from path-integrals to describe the ionic dynamics of a molecular
  junction in the presence of electrical current. The electronic
  environment serves as an effective non-equilibrium bath. The bath
  results in random forces describing Joule heating,
  current-induced forces including the non-conservative wind force,
  dissipative frictional forces, and an effective Lorentz-like force 
  due to the Berry phase of the non-equilibrium electrons. Using a generic 
  two-level molecular model, we highlight the importance of both current-induced
  forces and Joule heating for the stability of the system. We compare
  the impact of the different forces, and the wide-band approximation
  for the electronic structure on our result. We examine the current-induced
  instabilities (excitation of runaway ``waterwheel'' modes) and investigate
  the signature of these in the Raman signals.
\end{abstract}

\maketitle


\section{Introduction}
\label{sec:intro}
The interaction of electrons with local vibrations (phonons) has an important impact on the
conduction properties and stability of
molecular conductors\cite{ReZhMuBuTo97,SmNoUn.2002,RuAgVi96,YaBoBrAgRu98,OhKoTa98}
and has undergone intense study both experimentally and
theoretically\cite{PaLoSo.2003,Ta.2006,LiRa.2007,GaRaNi.2008,ScFrGa.2008,GaRaNi.2007,HoBoNeSaToFi06,StReHo98,AgYeRu.2003,AgUnRu.02,PaPaLiAnAlMc00,ViCuPaHa05,
FrBrLoJa.2004,PaFrBr.2005,FrPaBr.2007,PaFrUe.2008,SeRoGu.2005,PeRoDi.2007,McBoDu.2007,ChZwDi.2003,To.1998,SeNi.2002,
HuChDa.2007,NeShFi01,BrFl03,MiAlMi04,HaBeTh09,LoPeLaHo01,MiTiUe03,As08,luwang07,JiKuLuLu05,KuLaPaShSeBa04,GaNiRa06,TsTaKa.2008,VeStAl.2006,HiArRu.2008,
Smit.2004,Tsutsui.2006,Maria.2006,Ivan.2011,JoSe09,JoSe10}.
In the low bias regime where the voltage is comparable to phonon excitation energies, 
valuable information about the molecular conductor
can be deduced from the signature of electron-phonon interaction,
known as inelastic electron tunneling spectroscopy (IETS), or point
contact spectroscopy (PCS)\cite{SmNoUn.2002,AgUnRu.02,OkPaUe.2010,FoSALA.2011,KiPiEr.11}.
Theoretically, this regime has been addressed with some success using mean-field theory such as density
functional theory\cite{FrBrLoJa.2004,PaFrBr.2005,FrPaBr.2007,PaFrUe.2008,SeRoGu.2005,PeRoDi.2007,
JiKuLuLu05,OkPaUe.2010}, where the vibrations are assumed to be uncoupled to the electrons,
while the effect of the phonons on the electronic transport is taken into account using perturbation theory.
On the other hand, for higher voltage bias, and for highly transmitting systems, a large electronic current may
strongly influence the behavior of the phonons even for relatively weak electron-phonon coupling. The resultant
``Joule heating'' is well-known in the molecular electronics context\cite{To.1998,Ta.2006,GaRaNi.2007,HoBoNeSaToFi06}, 
and remains a lively area with a range of approaches (and with occasional lack of complete agreement 
between treatments \cite{McBoDu.2007,ChZwDi.2003}).

More recently the current-induced wind-force known from electromigration\cite{So.1998}
has been reexamined for atomic-scale conductors and shown also to be able to excite the
conductor and possibly lead to a runaway instability\cite{DuMcTo.2009,ToDuMc.2010,JMP.2010,ToDuDu.2011,LuGuBrHe.2011,BoKuEgVo.2011,BoKuEg.2012}.
It has been shown how a part of the force on an atom in the presence of the current may have a 
non-conservative (NC) component, able to do net work around closed paths.
This was explicitly proven by calculating the curl of the vector field describing the 
force on an atom\cite{DuMcTo.2009,ToDuDu.2011}. The NC energy transfer - also dubbed the 
atomic ``waterwheel'' effect - requires a generalized circular motion of the atoms
and involves the coupling of the electronic current to more than one vibrational mode.

Along with the NC force contribution we have recently identified a velocity-dependent 
current-induced force which conserves energy and acts as a Lorentz-like force
on the generalized circular motion. This force can be traced back to the
quantum mechanical Berry-phase (BP) of the electrons\cite{JMP.2010,BoKuEg.2012}.
Together with the NC force we will, further, have a component which is curl-free and is 
related to the change in the effective potential energy surface of the atoms due to 
the current\cite{DzKo.2011,NoPeMa.2011}, as well as nonadiabatic ``electronic friction'' 
forces\cite{HETU.1995}. In the nonequilibrium situation the ``friction'' force can, however, 
turn into a driving force amplifying the vibration. This happens under certain resonance 
conditions akin to a laser effect, but now involving phonons instead
of photons\cite{LuBrHe.2011,BoKuEgVo.2011}.

A unified approach including all aforementioned effects on an equal
footing is highly desirable for further study in this direction. In
this paper, we extend the electronic friction approach proposed by
Head-Gordon and Tully\cite{HETU.1995} for molecular dynamics to 
take into account the nonequilibrium nature of the electronic
current\cite{BrHe.1994a,BrHe95,MoHaMa.2006,JMP.2010,NoPeMa.2011,BoKuEg.2012}.
A similar approach has been taken to describe models of
nano-electro-mechanical systems (NEMS)\cite{HuMeZeBr.2010,MeBr.2011}.  
Using the Feynman-Vernon influence functional approach, we derive 
a semi-classical Langevin equation for the ions, which we can use to 
study Joule heating, current-induced forces, and heat transport in 
molecular conductors. We perform a perturbation expansion of the electron 
effective action over the electron-phonon interaction matrix. This 
allows us to make connections with other theoretical approaches, especially 
the nonequilibrium Green's function (NEGF) method, used to study the Joule
heating problem\cite{FrPaBr.2007,PeRoDi.2007}. We also give an extension 
of the perturbation result to the adiabatic limit, which makes connections 
with our previous results, and solves an infrared divergence problem in 
the expression of the BP force in Ref.~\onlinecite{JMP.2010}. We apply 
the theory to a two-level model in order to (1) clarify the roles played 
by different forces regarding the stability of the device, and (2) 
discuss the signature of the current-induced excitation in the Raman-scattering 
especially focussing on conditions close to a current-induced runaway instability.

The paper is organized as follows. In Sec.~\ref{sec:theory} we briefly review
the derivation of the generalized Langevin equation. In Sec.~\ref{sec:force} we
analyse the electronic forces entering into the Langevin equation.
Sec.~\ref{sec:model} compares the effect of different current-induced forces
for a two-level model, concentrating on the NC and BP force. In
Sec.~\ref{sec:extensions},  we extend the perturbation result to the adiabatic
limit, and introduce coupling of the system with electrode phonons.  The
derived formulas can be used to study the current-induced phononic heat
transport. In Sec.~\ref{sec:raman}, we present ways of calculating the quantum
displacement correlations, which is essential for the theoretical description
of Raman spectroscopy in the presence of current. Section~\ref{sec:conc} gives
concluding remarks.

\section{Theory}
\label{sec:theory}

\subsection{Influence functional theory}
\label{subsec:if}
We start from the influence functional theory of
Feynman and Vernon \cite{Feynman1963}, which treats the dynamics of a ``system'' in 
contact with a ``bath'' or ``reservoir''. In our case the system of interest 
consists of the few degrees of freedom describing the ions of a molecular conductor, 
interacting with the electronic reservoir composed of all electronic degrees of freedom 
in the molecule and electrodes, as well as the phonon reservoirs of two electrodes. 
The reservoirs can be out of equilibrium generating an electronic current,
and may, further, involve a temperature difference generating a heat flux 
between the electrodes. All effects of the bath are included in
the so-called influence functional, which gives an additional effective action,
modifying that of the isolated system. 

Now we briefly review the idea of the influence functional approach.
With the help of the influence functional, $F$, the reduced density matrix of the system
in the displacement representation reads,
\begin{eqnarray}
	\langle x_2|\rho_{s}(t_2)|y_2\rangle&=&\int dx_1 \int d y_1 \mathcal{K}(x_2,y_2;x_1,y_1)\nonumber\\
	&&\times \langle x_1|\rho_{s}(t_1)|y_1\rangle,
\label{eq:rhos}
\end{eqnarray}
with the propagator of the reduced density matrix
\begin{eqnarray}
	\mathcal{K}(x_2,y_2;x_1,y_1) &=& \int_{(x,y)(t_1)=(x_1,y_1)}^{(x,y)(t_2)=(x_2,y_2)}{\cal D}(x,y)\nonumber\\
	&\times& {\rm exp}\left[\frac{i}{\hbar}\left(S_{s}(x)-S_{s}(y)\right)\right]F(x,y).
\end{eqnarray}
Here $x$ and $y$ are a pair of displacement histories of the ions, and
$S_{s}$ is the action of the system only. In deriving this, we have assumed that 
the system and bath are uncorrelated at $t_1 (\to -\infty)$,
\begin{equation}
\rho(t_1)\:=\:\rho_{s}(t_1)\: \otimes \: \rho_{b}(t_1).
\end{equation}
The influence functional includes the information of the bath 
$S_b$ and its interaction with the system $S_i$,
\begin{eqnarray}
&&F(x,y)=\int d{r_2}d{r_1}d{q_1} \int_{(r, q)(t_1)=(r_1,q_1)}^{(r,q)(t_2)=(r_2,r_2)} {\cal D}(r,q)\nonumber\\
&& \times {\rm exp}\left[\frac{i}{\hbar}\left(S_{b}(r)+S_{i}(x,r)-S_{b}(q)-S_{i}(y,q)\right)\right]\nonumber\\
&&\times\langle r_1|\rho_{b}(t_1)|q_1\rangle.
\end{eqnarray}
with $r$ and $q$ representing forward and backward paths of the bath degrees of freedom. Most importantly, a correction to
the action of the system can be defined from the influence functional
$\Delta S=-i\hbar \, \ln F(x,y)$, which is usually not time-local or real.
It has been used to derive a semiclassical Langevin equation,
describing the dynamics of the system
interacting with the environment\cite{CALE.1983,SC.1982}.
In this case new variables are introduced,
\begin{eqnarray}
	Q = \frac{1}{2}(x+y),\quad\quad \xi = x-y\,,
	\label{eq:split}
\end{eqnarray}
describing the average and difference of the two paths,
respectively. In the semi-classical approach the average path, $Q$, is
shown to yield the variable in the Langevin equation, whereas role of
the difference, $\xi$, is to introduce fluctuating random forces in a
statistical interpretation. The influence of the environment will
favor paths with small excursions given by $\xi$, and will ensure
that only the solution, $Q$, obeying the classical path will
contribute for a high temperature reservoir. We illustrate this
further below.

Next, we introduce our model and give the result for the influence
functional describing the nonequilibrium electron bath. From the effective
action, we can read out the forces acting on the ions due to the electrons. 
We also discuss how a thermal flux may be included.
Parts of the derivations can be found in our previous
publications\cite{BrHe.1994a,BrHe95,JMP.2010}, but here we aim at 
a more general formulation, which we present with detailed derivations together with 
an illustrative model calculation. However we note that the theory is fully 
compatible with more realistic systems with complex electronic and vibrational
structure treated within a mean-field approach such as density functional theory.

\subsection{System setup and Hamiltonian}
\label{subsec:ham}

To obtain an effective action describing the vibrations (phonons) in the
molecular conductor we first divide the complete system into electron
and phonon subsystems. We will treat electron-electron interactions at the
mean-field level. To describe the nonequilibrium situation
where a current is flowing through the molecular conductor between two
reservoirs, the electron subsystem is further divided into a cental
part ($C$) and two electrodes ($L,R$), whose electrochemical potentials change
with applied bias. For the purposes of the present study, we allow electrons
to interact with phonons in $C$ only, and furthermore ignore the anharmonic 
coupling between these different modes. The coupling of the molecular
vibrations with electrode phonons will be considered in Sec.~\ref{sec:heat}.

The single particle mean-field electronic Hamiltonian at the relaxed ionic 
positions, ${H}^0$, is written within a tight-binding or LCAO type basis with 
corresponding electron creation (annihilation) operators for the $j$th orbital, 
$c^\dagger_j$ ($c_j$). The Hamiltonian $H^0$ spans the whole $LCR$ system, 
while the electron-phonon interaction is localized in $C$ \cite{FrPaBr.2007}.
The Hamiltonian of the whole system reads,
\begin{subequations}
	\begin{eqnarray}
		{H} &=& {H}_{ph} + {H}_e({u}),\label{equa}
		\\
		{H}_{ph} &=&\frac{1}{2} p^T p + \frac{1}{2}u^T K u,\label{equb}
		\\
		{H}_{e}({u})&=&\sum_{i,j}H^0_{ij} {c}^\dagger_i {c}_j+\sum_{k,i,j\in C}M^{k}_{ij}{c}^\dagger_i {c}_j {u}_k,\label{equc}
	\end{eqnarray}
\end{subequations}
where $p$ are momenta conjugate to $u$, and $u$ is a column vector containing 
the mass-normalized displacement operators of all ionic degrees of freedom
(e.g. $u_k=\sqrt{m_k}(r_k-r^0_k)$, where $m_k$ is the mass of ionic degree of 
freedom $k$ and $r_k$ ($r^0_k$) is its (equilibrium) position). 
The equilibrium zero-current dynamical matrix is denoted by $K$. This Hamiltonian 
has been used to describe IETS in molecular contacts with parameters obtained e.g. 
with DFT for concrete systems\cite{FrPaBr.2007,OkPaUe.2010}.
We have previously discussed the adiabatic limit, where the perturbation is 
in terms of the velocities $\dot u_k$ and where the full non-linear effects 
of $u_k$ can be included. Here we instead assume small displacements from 
equilibrium and expand the electronic Hamiltonian to first order in $u_k$. 
Later in Sec.~\ref{sec:adiabatic} we compare this to the adiabatic limit, 
discussed in Refs.~\onlinecite{JMP.2010,BoKuEg.2012}.

The correction to the action due to the coupling to the electron
reservoirs can be found using the linked-cluster expansion in the
coupling, $M^k$, following Ref.~\onlinecite{BrHe95}. The effective
action of the nonequilibrium, noninteracting electron bath reads,
\begin{eqnarray}
\Delta S(x,y)&=&i\hbar\sum_k \int_{0}^{1}d\lambda
\int_K d\tau\;\nonumber\\
&\times&{{\rm Tr}}[
{{G}}(\tau,\tau_+;{X}){M}^k{X}_k(\tau)]\;, \label{eq:seffff}
\end{eqnarray}
where the trace ${\rm Tr}\left[  \right]$ is over the electronic bath and this will be so in all following formulas.
The parameter $\lambda$ is used to keep track of the order in the linked-cluster expansion\cite{MahanBook}.
The time $\tau$ is defined on the Keldysh contour\cite{HaugJauhoBook} $K$. On the real time axis
the Green's function decomposes into
\begin{equation}
{{G}}(\tau,\tau')\:=\:\left ( \begin{array}{cc}
                 {G}(t,t')   & {G}^<(t,t')\\
                 {G}^>(t,t') & {\bar {G}}(t,t') \end{array} \right ),
\end{equation}
and
\begin{equation}
{X}(\tau)\:=\:\left (
\begin{array}{cc}
                  x(t) & 0\\
                 0    & -y(t)         \end{array} \right ).
\end{equation}
Time $\tau_+$ is infinitesimally later than $\tau$ on the whole Keldysh contour.
The limits of integration extend to $-\infty$ and $+\infty$ if not specified. 
This applies to all the integrals in the paper.
The Green's function is given by the Dyson equation,
\begin{eqnarray}
\label{eq:dyson}
&&{{G}}(\tau,\tau_+) = {{G}}_{0}(\tau,\tau_+)\\
&+& \sum_k\int_K d\tau'
{{G}}_{0}(\tau,\tau'){{M}^k}{{X_k}}(\tau')
{{G}}(\tau',\tau_+), \nonumber
\end{eqnarray}
${{G}}_{0}$ being the single electron Green's function without interaction with the ions, which reads
\begin{eqnarray}
&&{{G}}_{0}(\tau,\tau')=i\sum_\alpha \int_{}^{}\frac{d\varepsilon}{2\pi\hbar}\;
e^{-\frac{i}{\hbar}\varepsilon(t-t')}A_\alpha(\varepsilon)\times\\
\!\!\!\!&&\left ( \begin{array}{cc}
   n_F(\varepsilon-\mu_\alpha)-\theta(t-t') & n_F(\varepsilon-\mu_\alpha)\\
   n_F(\varepsilon-\mu_\alpha)-1  & n_F(\varepsilon-\mu_\alpha)-\theta(t'-t) \end{array}
   \right )\nonumber.
\end{eqnarray}
$n_F(\varepsilon-\mu_\alpha)=1/\left[1+e^{(\varepsilon-\mu_\alpha)/k_BT}\right]$
is the Fermi-Dirac distribution function for electrode $\alpha$, $A=\sum_\alpha A_\alpha$, and
$\theta$ is the Heaviside step function. The spectral function is defined as
\begin{equation}
	A_\alpha(\varepsilon)= iG_0^r(\varepsilon)[\Sigma^r_\alpha(\varepsilon)-\Sigma_\alpha^a(\varepsilon)]G_0^a(\varepsilon),
\end{equation}
and $\Sigma^r_\alpha(\varepsilon)$ ($\Sigma^a_\alpha(\varepsilon)$)
the retarded (advanced) electron self-energy from lead $\alpha$. We
use $\varepsilon$ and $\omega$ as parameters for the electron and
electron-hole pair/phonon properties, respectively.

The effective action in \Eqref{eq:seffff} could be expanded into 
an infinite series in $M$. We only keep terms up to second order,
assuming small $M$ (or alternatively assuming small displacements 
as stipulated earlier). The first-order contribution written in terms of the
average and difference paths for the vibrations, $Q_k,\xi_k$, reads,
\begin{eqnarray}
	\Delta S^{(1)}(Q,\xi)&=& \sum_kF^{(1)}_{k}\:\int_{}^{}dt\;\xi_k(t),
	\label{eq:1st}
\end{eqnarray}
with a first-order, displacement-independent force term,
\begin{eqnarray}
  &&F^{(1)}_{k} = - 2\sum_{\alpha} \int\frac{d\varepsilon}{2\pi}\;
  {{\rm Tr}}[A_\alpha(\varepsilon)M^k]\Delta n^\alpha_F(\varepsilon), \label{eq:fek0}\\
  &&\Delta n^\alpha_F(\varepsilon) = n_F(\varepsilon -\mu_\alpha)-n_F(\varepsilon-\mu_0).
	\label{eq:fek}
\end{eqnarray}
Here $\mu_0$ is the equilibrium electrochemical potential and the 
factor of $2$ accounts for spin degeneracy. Above we explicitly
subtract the equilibrium forces (obtained for $\mu_\alpha=\mu_0$), 
since these are already included in the elastic forces described by
$K$. The different filling of electronic states
originating from different electrodes in Eq.~(\ref{eq:fek}) results in a 
displacement-independent ``wind'' force\cite{So.1998,BrStTa.2003,BeGuWiZh.2010}. 
Its effect amounts in the small-displacement 
approximation used here to a bias-induced shift of the equilibrium ionic positions. 
We will therefore ignore it from now on, since we will be considering the 
effects of the nonequilibrium electrons on the ionic dynamics.

Before introducing the second-order contribution, we note that the applied
bias between the two electrodes also modifies the electronic
Hamiltonian, and thus $A_\alpha$ and $M^k$. It is the result of charge
rearrangement in the device in response to the applied bias. This is out of
the scope of present paper, and is not included in Eq.~(\ref{eq:fek}).  The
inclusion of external electric fields in the electronic Hamiltonian can
account, for example, for the ``direct'' electromigration
force\cite{So.1998}. 


We now turn to the second-order contribution central to our discussion,
\begin{eqnarray}
\label{eq:s2}
&&\Delta S^{(2)}(Q,\xi)=-\frac{i}{4}\sum_{\alpha,\beta,l,k}\int dt\int dt'
\;\int_{}^{}d\omega\;\nonumber\\
&\times&e^{-i\omega(t-t')}\:\Lambda_{kl}^{\alpha\beta}(\omega)
\Bigg[\coth\left(\frac{\hbar\omega-(\mu_\alpha-\mu_\beta)}{2k_BT}\right)\nonumber\\
&\times&\xi_k(t)\xi_l(t')
+2\theta(t-t')\xi_k(t)Q_l(t')\nonumber\\
&-&2\theta(t'-t)\xi_l(t')Q_k(t) \Bigg],
\end{eqnarray}
where the central quantity is an interaction-weighted electron-hole pair density of states (incl. spin),
\begin{eqnarray}
\label{eq:ehdos}
\Lambda^{\alpha\beta}_{kl}(\omega)&=&2\int_{}^{}\frac{d\varepsilon_1}{2\pi}\;
\int_{}^{}\frac{d\varepsilon_2}{2\pi}\;
\delta(\hbar\omega-\varepsilon_1+\varepsilon_2)\;\nonumber\\
&\times&{ {\rm Tr}}[M^kA_\alpha(\varepsilon_1)M^lA_\beta(\varepsilon_2)]\\
&\times&(n_F(\varepsilon_1-\mu_\alpha)-n_F(\varepsilon_2-\mu_\beta))\nonumber.
\end{eqnarray}
It has the following properties,
\begin{eqnarray}
&&\Lambda^{\alpha\beta}_{kl}(\omega)={\Lambda^{\alpha\beta}_{lk}}^*(\omega), \label{eq:p1}\\
&&\Lambda^{\alpha\beta}_{kl}(\omega)=-\Lambda^{\beta\alpha}_{lk}(-\omega),
	\label{eq:p2}\\
&&\Lambda(-\omega)=-\Lambda^*(\omega),
\end{eqnarray}
where we have defined $\Lambda=\sum_{\alpha,\beta}\Lambda^{\alpha\beta}$.

\subsection{The generalised Langevin equation}
\label{subsec:gl}
In order not to complicate the equations we will in the following suppress the
phonon-mode index and implicitly write vectors and matrices without these. 
Note that  these phonon indices are generally not interchangeable. This calls
for care for example when carrying out permutations within the trace in
$\Lambda^{\alpha\beta}_{kl}$ and quantities derived from it. If the reader wishes 
to make such rearrangements, it is necessary to reinstate the indices ($k$, $l$ 
above), in the correct starting order, first. Using the results
of Subsec.~\ref{subsec:if} we get the path-integral,
\begin{eqnarray}\label{eq:Z1}
	\mathcal{K}&=&\int \mathcal{D}\xi \int \mathcal{D}Q \exp\left[-\frac{i}{\hbar}\int dt \int dt' \xi^T(t)L(t,t')Q(t')\right]\nonumber\\
	&\times& \exp\left[-\frac{1}{2\hbar}\int dt \int dt'\xi^T(t)\hat{\Pi}(t-t')\xi(t')\right].
	\label{eq:tact}
\end{eqnarray}
We have defined
\begin{eqnarray}\label{eq:Z2}
	&&L(t,t') = \left(\frac{\partial^2}{\partial t^2}+K\right)\delta(t-t')+{\Pi}^r(t-t'),\\
	&&\tilde{\Pi}(t-t') = 2\pi i \, \mathcal{F}^{-1}\{\Lambda(\omega)\},\\
	&&{\Pi}^r(t-t') = \theta(t-t')\tilde{\Pi}(t-t'),
\end{eqnarray}
and
\begin{eqnarray}
	&&\hat{\Pi}(t-t')=\mathcal{F}^{-1}\{\hat{\Pi}(\omega)\},
	\label{eq:lo}
\end{eqnarray}
with
\begin{eqnarray}
	&&\hat{\Pi}(\omega)\equiv \hat{\Pi}_0(\omega)+\Delta\hat{\Pi}(\omega)\nonumber\\
	&&=-\pi\Lambda(\omega) \coth\left(\frac{\hbar\omega}{2k_BT}\right)-
	\pi\sum_{\alpha\beta}\Lambda^{\alpha\beta}(\omega)\nonumber\\
	 &&\times\left[\coth\left(\frac{\hbar\omega-(\mu_\alpha-\mu_\beta)}{2k_BT}\right)-\coth\left(\frac{\hbar\omega}{2k_BT}\right)\right].
\end{eqnarray}
We have split $\hat{\Pi}$ into two terms. We will see in Sec.~\ref{sec:heat} that $\Delta\hat{\Pi}$ is 
responsible for the heating effect. The Fourier transform is defined as $\mathcal{F}\{f(t)\} = \int dt f(t) e^{i\omega t}$,
and $\mathcal{F}^{-1}\{f(\omega)\} = \int \frac{d\omega}{2\pi} f(\omega) e^{-i\omega t}$.
After a Hubbard-Stratonovich transformation, we get
\begin{eqnarray}
	\mathcal{K}&=&\int \mathcal{D}\xi \int \mathcal{D}Q \int \mathcal{D}f \nonumber\\
	&\times& 
	\exp \left(\frac{i \xi^T_2 \dot Q_2}{\hbar} \right) \exp \left(- \frac{i \xi^T_1 \dot Q_1}{\hbar} \right) \nonumber\\
	&\times&\exp\left[-\frac{i}{\hbar}\int dt \, \xi^T(t)\left(\int dt'L(t,t')Q(t')-f(t)\right)\right]\nonumber\\
	&\times&\exp\left[-\frac{1}{2\hbar}\int dt \int dt' f^T(t)\hat{\Pi}^{-1}(t,t')f(t')\right].
	\label{eq:tact2}
\end{eqnarray}
The factors in line 2 (where $\xi_1 = \xi(t_1)$, etc) arise from the integration by parts 
to transform the kinetic energy ($\dot \xi^T(t) \dot Q(t)$) into the form in line 3, 
and are needed for example if one wishes to make a connection with the Wigner function.
The above form of the effective action suggests a classical interpretation to the
motion of the average displacement $Q$. It follows a generalized Langevin
equation\cite{CALE.1983,SC.1982}
\begin{eqnarray}
	\ddot Q(t)\!=\!-K Q(t)-\!\int{\Pi}^r(t-t')Q(t')dt' + f(t),
	\label{eq:langa}
\end{eqnarray}
where $f(t)$ is a classical stochastic force, whose time-correlation is given by 
\begin{equation}
	\langle f(t)f^T(t')\rangle=\hbar\hat{\Pi}(t-t').
\end{equation}
Equation (\ref{eq:langa}) is the equation of motion for harmonic oscillators,
perturbed by the second and third terms due to interaction with electrons.
There are two ways of seeing how it arises. First, it is the Euler-Lagrange equation 
of motion for the action in Eq. (\ref{eq:tact2}). Alternatively, if in (\ref{eq:tact2}) 
we carry out the $\xi$-integral, that introduces $\delta( L \cdot Q - f )$ (where $\cdot$ denotes 
time-integration for short) in the integrand, which restricts $Q(t)$ to the evolutionary path
generated by (\ref{eq:langa}). The physical significance of $Q(t)$ is that it is the quasi-classical
coordinate appearing in the Wigner function. Its Newtonian equation of motion above relies on the 
quadratic nature of the effective action as a functional of the generalised coordinates, and would have
to be revisited in the presence of higher-order terms in the Hamiltonian, or in the Green's function expansion.

It is possible to solve the generalised Langevin equation by Fourier transform. 
From that, we get the semi-classical displacement correlation function,
\begin{equation}
	\frac{1}{\hbar}\,\langle Q Q^T \rangle (\omega) = D^r(\omega) \hat{\Pi}(\omega) D^a(\omega).
        \label{eq:qqq}
\end{equation}
Here the phonon retarded Green's function is defined as
\begin{equation}
	D^r(\omega) =  (D^a(\omega))^\dagger = \left [ (\omega+i0^+)^2 - K - \Pi^r(\omega)\right]^{-1} .
	\label{eq:dr}
\end{equation}
Note that $\tilde{\Pi}$  and $\hat{\Pi}$ can be written as the standard phonon self-energies in the NEGF method
\begin{eqnarray}
&&\tilde{\Pi}(\omega)=\Pi^r(\omega)-\Pi^a(\omega)=\Pi^>(\omega)-\Pi^<(\omega),\\
&&\hat{\Pi}(\omega)=\frac{i}{2}\left[\Pi^>(\omega)+\Pi^<(\omega)\right].
	\label{eq:pp1}
\end{eqnarray}
The self-energy diagram is shown in \Figref{fig:bubble}, and their expressions are given in
Appendix~\ref{sec:se}.

\begin{figure}[htpb]
	\begin{center}
		\includegraphics[scale=2.5]{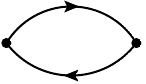}
	\end{center}
	\caption{Lowest order phonon self-energy due to interaction with electrons.}
	\label{fig:bubble}
\end{figure}

\section{Forces}
\label{sec:force}
\subsection{General results}
\label{subsec:gr}
The electronic forces in the Langevin equation are divided into
stochastic and deterministic parts. The correlation function of
the stochastic force has the following properties: 
\begin{subequations}
	\begin{eqnarray}
		\hat{\Pi}^\dagger(\omega) &=& \hat{\Pi}(\omega),\\
		\hat{\Pi}(-\omega) &=& \hat{\Pi}^*(\omega).
	\end{eqnarray}
\end{subequations}
Consequently, $\hat{\Pi}(t)$ is real, but in general $\hat{\Pi}(-t) \neq
\hat{\Pi}(t)$ at finite bias.

Now let us look at the deterministic forces due to electrons. 
$\tilde{\Pi}(\omega)$ has the properties: 
\begin{subequations}
	\begin{eqnarray}
	\tilde{\Pi}^\dagger(\omega) = -\tilde{\Pi}(\omega),\\
	\tilde{\Pi}(-\omega) = \tilde{\Pi}^*(\omega).
	\end{eqnarray}
\end{subequations}
In the Fourier domain, we can split it into different contributions,
\begin{eqnarray}
	\label{eq:tll}
	&&-\Pi^r(\omega)  =-i\pi{\rm Re}\Lambda(\omega) +\pi{\rm Im}\Lambda(\omega) \nonumber\\
	&&-\pi\mathcal{H}\{{\rm Re}\Lambda(\omega')\}(\omega)  -i\pi\mathcal{H}\{{\rm Im}\Lambda(\omega')\}(\omega) ,
\end{eqnarray}
where the Hilbert transform is defined as $	\mathcal{H}\{g(x')\}(x) = \frac{1}{\pi}\mathcal{P} \int \frac{g(x')}{x'-x}dx'$.
Using the symmetry properties of $\Lambda(\omega)$, we can now examine each term in
\Eqref{eq:tll}.

The first term is imaginary and symmetric. It describes the
standard friction, i.e. processes whereby the motion of vibrating ions
generate electron-hole pairs in the electronic environment. 
This process exists also in equilibrium. We can write
\begin{equation}
	F_{FR}=-\eta(\omega)\dot{Q}(\omega),
	\label{}
\end{equation}
with the friction matrix defined by,
\begin{equation}
	\eta(\omega) = -\frac{\pi}{\omega}{\rm Re}\Lambda(\omega).
	\label{eq:gfr}
\end{equation}

The second term in \Eqref{eq:tll} is real and anti-symmetric.  It has a finite value 
even in the limit of zero frequency. It is describing the NC force, discussed
recently\cite{DuMcTo.2009,JMP.2010,ToDuMc.2010,BoKuEgVo.2011,ToDuDu.2011,BoKuEg.2012}
\begin{equation}
	F_{NC} = \mathcal{N}(\omega)Q(\omega),
	\label{}
\end{equation}
with
\begin{equation}
	\mathcal{N}(\omega)=\pi {\rm Im}\Lambda(\omega).
	\label{}
\end{equation}

The third term is real and symmetric, and can be considered a 
renormalization (RN) of the dynamical matrix
\begin{equation}
	F_{RN} = -\zeta(\omega)Q(\omega),
	\label{}
\end{equation}
with
\begin{equation}
	\zeta(\omega)=\pi\mathcal{H}\{{\rm Re}\Lambda(\omega')\}(\omega).
	\label{}
\end{equation}

Finally, the last term is imaginary and anti-symmetric, proportional to
$\omega$ for small frequencies.  Hence it is to be identified with the BP force
in Ref.~\onlinecite{JMP.2010},
\begin{equation}
	F_{BP} = -\mathcal{B}(\omega)\dot{Q}(\omega),
	\label{}
\end{equation}
with the effective magnetic field
\begin{equation}
	\mathcal{B}(\omega) =-\frac{\pi}{\omega}\mathcal{H}\{{\rm Im}\Lambda(\omega')\}(\omega).
	\label{eq:emag}
\end{equation}

\subsection{Equilibrium and nonequilibrium contributions}
\label{subsec:wb0}
We can divide the $\Lambda(\omega)$ into an equilibrium part and a
nonequilibrium part
$\Lambda(\omega)=\Lambda_{eq}(\omega)+\Delta\Lambda(\omega)$, and look
at their contribution to the forces separately.
$\Lambda_{eq}(\omega)$ is given by \Eqref{eq:ehdos} with $\mu_\alpha$
and $\mu_\beta$ replaced by the equilibrium electrochemical potential
$\mu_0$. The nonequilibrium part can be written as
\begin{eqnarray}
 \label{eq:dl}
  &&\left(\begin{array}{cc}{\rm Im}\\{\rm Re}\end{array}\right)\Delta\Lambda(\omega)=2\sum_\alpha\int\frac{d\varepsilon}{4\pi^2}\Delta n_F^\alpha(\varepsilon)\\
  &&\times\left(\begin{array}{cc}{\rm Im}\\{\rm Re}\end{array}\right){\rm Tr}\left[ MA_\alpha(\varepsilon)M\left(A(\varepsilon_-)\left(\begin{array}{cc}+\\-\end{array}\right) A(\varepsilon_+)\right) \right],\nonumber
\end{eqnarray}
with $\varepsilon_\pm=\varepsilon\pm\hbar\omega$.

In the following we assume zero magnetic fields and treat $M^k$ and $A(\epsilon)$ 
as real symmetric matrices in the electronic real-space basis.

\subsubsection{Equilibrium contribution}
We consider first the equilibrium part $\Lambda_{eq}(\omega)$. It
is real, giving the equilibrium friction, and its Hilbert transform
gives the equilibrium renormalization of the potential.

\emph{Friction --} The equilibrium friction matrix reads
\begin{eqnarray}
  &&\eta_{eq} (\omega)=2\frac{1}{2\omega}\int \frac{d\varepsilon}{2\pi}n_F(\varepsilon-\mu_0)\nonumber\\
  &\times&{\rm Tr}\left[M A(\varepsilon)M\left(A(\varepsilon_+)-A(\varepsilon_-)\right)\right].  \label{eq:frr}
\end{eqnarray}

\emph{Renormalization -- } The equilibrium RN reads
\begin{eqnarray}
	\zeta_{eq}(\omega)&=&2\int\frac{d\varepsilon}{2\pi}n_F(\varepsilon-\mu_0)\\
	&\times&{\rm Tr}\left[M A(\varepsilon)M\left( R(\varepsilon_-)+R(\varepsilon_+) \right)\right]\nonumber.
	\label{eq:ern0}
\end{eqnarray}
We have defined 
\begin{equation}
R(\varepsilon) = - \frac{1}{2}\,\mathcal{H}\{A(\varepsilon')\}(\varepsilon) = \frac{G_0^r(\varepsilon) + G_0^a(\varepsilon)}{2}. 
\end{equation}
In general $\zeta_{eq}(\omega)$ has a frequency dependence, of ${\cal O}(\omega^2)$ or higher.
Its static (frequency-independent) part is already included in the dynamical matrix, when 
calculated within the Born-Oppenheimer approximation.

\subsubsection{Nonequilibrium contribution}
Now we consider contributions from the the nonequilibrium part
$\Delta\Lambda(\omega)$. Although we are considering the two-terminal 
$LCR$ assembly described earlier, an arbitrary number of independent terminals 
can be accommodated via the summations over indices $\alpha$, $\beta$.
In the two-terminal case, we write $eV=\mu_L-\mu_R$ where $V$ is the bias.

\emph{Friction --}We first get a
correction to the equilibrium friction
\begin{eqnarray}
&&\Delta\eta(\omega)=2\sum_\alpha\int \frac{d\varepsilon}{2\pi}\frac{\Delta n^\alpha_F(\varepsilon)}{2\omega}\\
	&\times&{\rm Re}{\rm Tr}\left[ M(A(\varepsilon_+)-A(\varepsilon_-))MA_\alpha(\varepsilon) \right].\nonumber
	\label{eq:nfr}
\end{eqnarray}
This nonequilibrium correction may give rise to another interesting instability, 
characterized by a negative friction, if the spectral functions depend on the energy in a
special way which enable a population-inverted situation\cite{LuBrHe.2011}. It is responsible
also for enhanced heating, and for the converse: current-facilitated thermal relaxation, 
in systems with appropriate spectral features \cite{heating.2008,cooling.2009}.

\emph{NC force --} The coefficient for the NC force is
\begin{eqnarray}
	&&\mathcal{N}(\omega)
	=2\sum_\alpha \int\frac{d\varepsilon}{2\pi}\frac{\Delta n_F^\alpha(\varepsilon)}{2}\\
	&\times& {\rm Im}{\rm Tr}\left[ M A_\alpha(\varepsilon) M(A(\varepsilon_+)+A(\varepsilon_-)) \right].\nonumber
	\label{eq:dnc}
\end{eqnarray}
Performing the Hilbert transform, we get the nonequilibrium correction to the RN force and the BP force.

\emph{Renormalization --} The nonequilibrium correction to the RN force is given by the coefficient
\begin{eqnarray}
	&&\Delta\zeta^{}(\omega)=2\sum_\alpha \int\frac{d\varepsilon}{2\pi}\Delta n_F^\alpha(\varepsilon)\nonumber\\
	&\times& {\rm Re}{\rm Tr}\left[ MA_\alpha(\varepsilon) M(R(\varepsilon_+)+R(\varepsilon_-))\right].
	\label{eq:z1}
\end{eqnarray}

\emph{BP force --} The BP force is
\begin{equation}
	F_{BP}(\omega) =-\mathcal{B}(\omega)\dot{Q}(\omega),
	\label{eq:nbp}
\end{equation}
with
\begin{eqnarray}
&&\mathcal{B}(\omega)=2\sum_\alpha \int \frac{d\varepsilon}{2\pi}\frac{\Delta n_F^\alpha(\varepsilon)}{\omega}\nonumber\\
&\times&{\rm Im}{\rm Tr}\left[ M A_\alpha(\varepsilon) M(R(\varepsilon_+)-R(\varepsilon_-))\right].
	\label{eq:z2}
\end{eqnarray}
The relative magnitude of the BP force and NC force can be estimated as,
\begin{equation}
	\frac{|F_{BP}|}{|F_{NC}|} \sim \frac{\hbar\omega}{|H|},
	\label{eq:bpnc}
\end{equation}
$|H|$ being a typical electronic hopping integral.
Ref.~\onlinecite{JMP.2010}, instead of $|H|$, used the phonon frequency as
a cutoff when calculating the BP force. This over-estimates the effect of the
BP force. A more detailed discussion of this is given in Appendix \ref{sec:nl}.

If the dynamics of the ions is very slow compared to the dynamics of the electrons,
the electronic spectrum varies weakly within the vibrational energy spectrum, and
we can take the $\omega \to 0$ limit in the expressions for the forces. All deterministic
forces then become time-local. This will be compared with the adiabatic result in Sec.~\ref{sec:adiabatic}.

\subsection{Wideband approximation}
\label{subsec:wb1}

A regime of practical interest is the limit where the electronic spectrum varies 
slowly not only within the vibrational energy spectrum, but also within the bias window. 
Then we can ignore its energy dependence altogether and
evaluate all electronic properties at the Fermi level ($\mu_0$). This is
the wideband approximation used for example in Ref.~\onlinecite{PaFrBr.2005} to study IETS.

\subsubsection{Dynamical equations}

Within the above approximation, the Langevin equation reads
\begin{eqnarray}
	\label{eq:lang}
	\ddot Q(t) &=& -KQ(t)-\eta_0 \dot{Q}(t) + \mathcal{N}_0 Q(t)\nonumber\\
	&-&\zeta_0^{} Q(t) - \mathcal{B}_0^{} \dot{Q}(t)+ f_{}(t),
\end{eqnarray}
with
\begin{eqnarray}
	\eta_0&=&2\frac{\hbar}{4\pi}{\rm Tr}\left[MA(\mu_0)MA(\mu_0)\right],\\
	\mathcal{N}_0&=&eV\chi^{-}, \label{eq:defco0}\\
	\zeta^{}_0&=&2\frac{eV}{2\pi}{\rm Re}{\rm Tr}\left[M \Delta A(\mu_0) M R(\mu_0)\right],\\
	\mathcal{B}^{}_0&=&2\frac{\hbar eV}{2\pi}{\rm Im}{\rm Tr}\left[M \Delta A(\mu_0) M \partial_\varepsilon R(\mu_0)\right],
	\label{eq:defco}
\end{eqnarray}
where we have introduced,
\begin{eqnarray}
	\chi_{}^{-}&=&2\frac{1}{2\pi}{\rm Im}\textrm{Tr}[MA_L(\mu_0)MA_R(\mu_0)],\\
	\chi_{}^{+}&=&2\frac{1}{2\pi}{\rm Re}\textrm{Tr}[MA_L(\mu_0)MA_R(\mu_0)],
	\label{eq:defco1}
\end{eqnarray}
and $\Delta A(\mu_0)=A_L(\mu_0)-A_R(\mu_0)$.
The noise correlation function also takes a simpler form,
\begin{equation}
	\label{eq:oeq}
	\hat{\Pi}_{0}(\omega)=(\omega\eta_0 - i eV\chi^-) \coth\left(\frac{\hbar\omega}{2k_BT}\right),
\end{equation}
and
\begin{eqnarray}
	\label{eq:onon}
	 \Delta\hat{\Pi}_{}(\omega)&=&\frac{1}{2}\sum_{\sigma=\pm}\left(\chi^+ -i \sigma \chi^-\right)\left(\hbar\omega +\sigma eV\right)\\
	&\times&\left[\coth\left(\frac{\hbar\omega +\sigma eV}{2k_BT}\right)-\coth\left(\frac{\hbar\omega}{2k_BT}\right)\right].\nonumber
\end{eqnarray}
The noise originates from the fluctuating part of the forces (\Eqref{eq:xi}),
whose correlation spectrum in frequency space is in general Hermitian, but not
real at finite bias. The real part corresponds to the friction, and the
imaginary part to the NC and BP force in the deterministic forces. Importantly,
the quantum zero-point fluctuation is taken into account (encoded in the
$\coth$ function).

\subsubsection{Phonon excitation}
\begin{figure}[htpb]
	\begin{center}
		\includegraphics[scale=0.7]{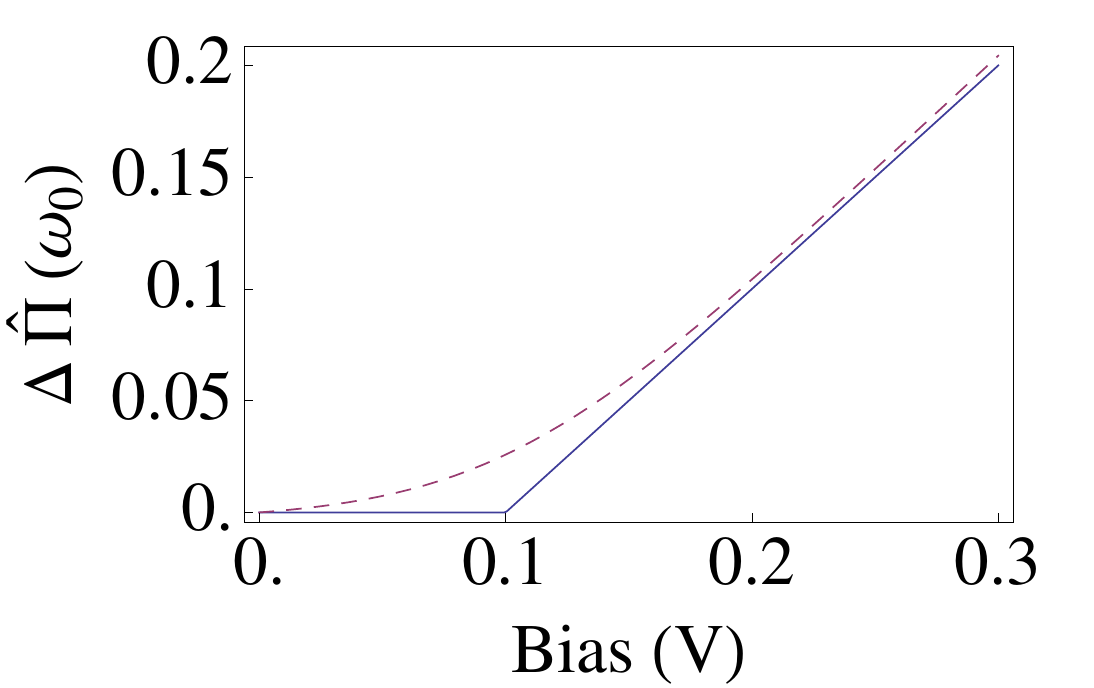}
	\end{center}
	\caption{Example of the nonequilibrium noise spectrum at $T = 0$ K (solid)
and $300$ K (dashed) for a given phonon mode with frequency $\hbar\omega_0=0.1$ eV.}
	\label{fig:non}
\end{figure}

When isolated from the electrode phonons, the system could get heated or cooled
due to the passing electrical current. At steady state, the system phonon
population will be different from that at equilibrium. We will now compare the 
phonon excitation result from the Langevin equation with that from NEGF theory\cite{FrPaBr.2007}. 
To do that, we employ the wideband approximation, and ignore couplings between 
different phonon modes. In this way we can study each mode separately. At steady state, 
the energy stored in each phonon mode can be obtained from the solution for \Eqref{eq:lang} 
in frequency space,
\begin{eqnarray}
 	\label{eq:ei}
	E_i &=& \langle \dot{Q}^2_i(t) \rangle  \nonumber\\
	&=& \int_{}^{} \omega^2 \langle Q_i Q_i\rangle (\omega)\frac{d\omega}{2\pi}.
\end{eqnarray}
Using \Eqref{eq:qqq} and Eqs.~(\ref{eq:oeq}-\ref{eq:onon}), assuming small broadening of the phonon mode, it can be further simplified as
\begin{eqnarray}
	 E_i&\approx&\frac{\hbar\omega_i}{2}\coth\left(\frac{\hbar\omega_i}{2k_BT}\right)+\frac{\hbar\Delta\hat{\Pi}_{ii}(\omega_i)}{2\eta_{ii}}\nonumber\\
	&=& \left(N_i+\frac{1}{2}\right)\hbar\omega_i.
	\label{eq:ploe}
\end{eqnarray}
Here we have introduced an effective phonon number $N_i$.
At low temperature $\Delta\hat{\Pi}_{ii}(\omega)$ can be approximated as,
\begin{equation}
	\Delta\hat{\Pi}_{ii}(\omega>0) \propto (eV-\hbar\omega)\theta(eV-\hbar\omega).
	\label{eq:non2}
\end{equation}
Figure~\ref{fig:non} shows the bias dependence of
$\Delta\hat{\Pi}_{ii}$ at zero and finite temperature for a phonon mode
$\hbar\omega_i=0.1$ eV. Interestingly, the Joule heating exhibits 
a threshold for phonon excitation at the
phonon energy, at zero temperature. If fact, \Eqref{eq:ploe} is 
exactly the same as the quantum result Eq.~($47$) in
Ref.~\onlinecite{FrPaBr.2007}. If we take the different
definition of the electron-phonon interaction matrix used here 
and in Ref.~\onlinecite{FrPaBr.2007}, we find that the friction
coefficient $\eta_{ii}$, and the nonequilibrium noise spectrum
$\Delta\hat{\Pi}_{ii}(\omega_i)$ are related to the electron-hole pair
damping $\gamma_{e-h}^i$, and phonon emission rate $\gamma_{em}^i$
defined in Ref.~\onlinecite{FrPaBr.2007} as $\gamma_{e-h}^i =
\eta_{ii}$, $\Delta\hat{\Pi}_{ii}(\omega_i) =
2\omega_i\gamma_{em}^i$. We therefore conclude that we recover the
quantum-mechanical result from the semi-classical Langevin
equation. This is because we only need the quantum average of the
equal-time displacements $\langle u_i(t)u_j(t)\rangle$ to study the
energy transport which can be calculated exactly from the semi-classical
Langevin equation (see Appendix~\ref{app:qccor} for details).
Alternatively, in Wigner-function language, the mean phonon energy is expressible 
solely in terms of the coordinate $Q$ (and/or the velocity $\dot Q$). As we have
seen, for a harmonic action the present Newtonian equation of motion for $Q$ is exact.

\section{Numerical results for a simple two-level model}
\label{sec:model}
\begin{figure}[htpb]
	\begin{center}
		\includegraphics[scale=0.5,angle=0]{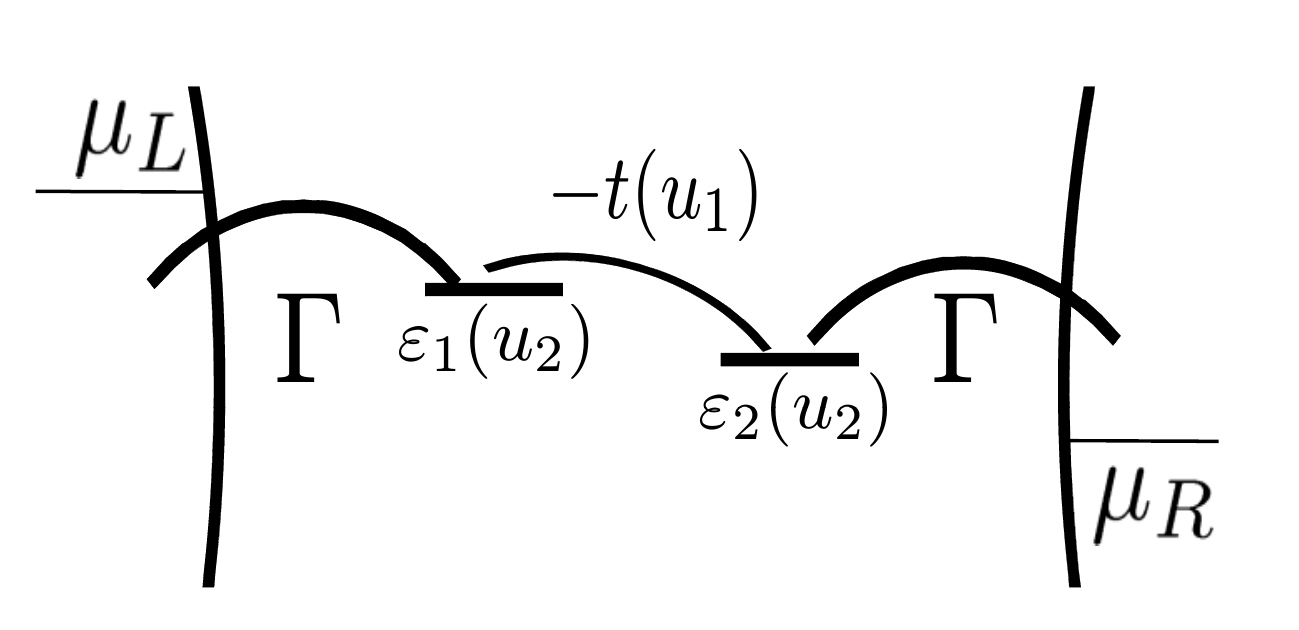}
	\end{center}
	\caption{Schematic diagram of the two-level model.
The bias is defined as $eV=\mu_L-\mu_R$ and the average electrochemical potential $\mu_0=(\mu_L+\mu_R)/2$.}
	\label{fig:model}
\end{figure}

To build an intuitive understanding of the theory above,
we now apply it to a simple spinless two-level model which 
could describe a diatomic molecule. For this model system, 
it is possible to do the calculation using the general 
results in Sec.~\ref{subsec:gr}, without the
approximations developed in Sec.~\ref{subsec:wb1}.
We start from a model Hamiltonian for the isolated system,
\begin{eqnarray}
	H &=& H_e+H_{ph}+H_{int}\nonumber\\
	&=& \varepsilon_0 (c^\dagger_1 c_1-c^\dagger_2 c_2)-t(c^\dagger_1c_2+c^\dagger_2c_1)\nonumber\\
	&+& \sum_{i=1,2}\left(\frac{1}{2}\dot{u}_i^2+\frac{1}{2}\omega_i^2u_i^2+H^{i}_{int}\right).
	\label{eq:ham}
\end{eqnarray}
The electrons couple with two phonon modes in the following two forms:
\begin{equation}
	H^{1}_{int}=m_1 u_1(c^\dagger_1c_2+c^\dagger_2c_1),
	\label{eq:int1}
\end{equation}
and
\begin{equation}
	H^{2}_{int}=m_2 u_2(c^\dagger_1c_1-c^\dagger_2c_2).
	\label{eq:int2}
\end{equation}
The first and second electronic levels couple with the left and right electrode, respectively,
with level broadening $\Gamma$. A sketch of the model is shown in Fig.~\ref{fig:model}.
Mode 1 corresponds to the bond-stretching mode of the diatomic molecule, while mode 2 mimics 
the rigid motion of the diatomic molecule between the two electrodes.

\subsection{Current-induced forces and phonon excitation}
\begin{figure}[htpb]
	\begin{center}
		\includegraphics[scale=0.5]{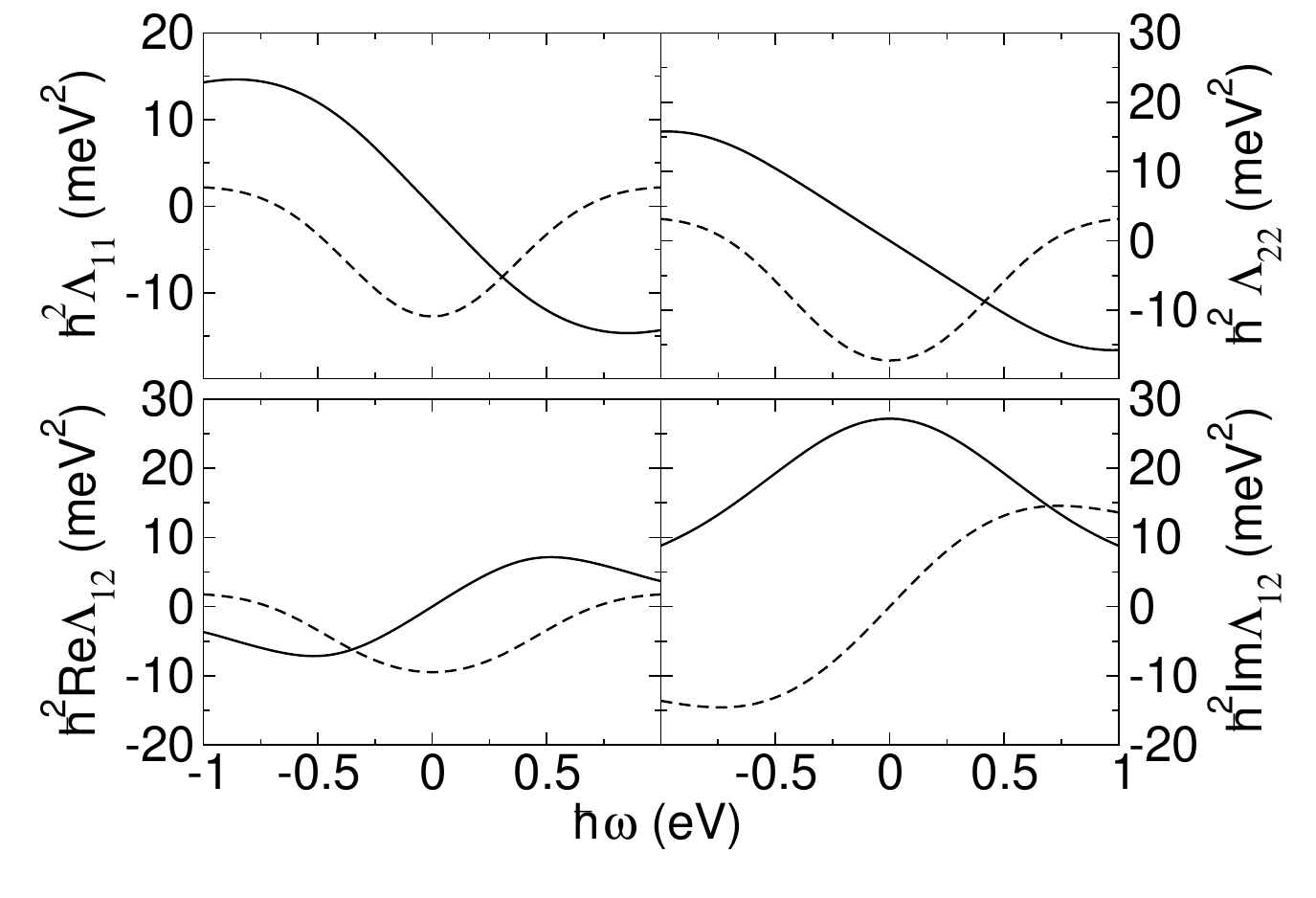}
	\end{center}
	\caption{Different parts of the $\Lambda(\omega)$ function (solid) and
	their Hilbert transform (dashed). The model parameters are $\Gamma=1$ eV, $t=0.2$ eV,
$\varepsilon_0=0$, $m_1=m_2=0.01$ eV/$\sqrt{{\rm amu}}$\AA, $\mu_0 = 0$, and $V=1$ V. 
Indices 1 and 2 refer to the respective phonon modes. The equilibrium renormalization term has
	already been substracted in the plot. We use the same parameters in the
	following figures if not stated explicitly.}
	\label{fig:lam}
\end{figure}

Figure~\ref{fig:lam} shows different parts of $\Lambda(\omega)$ and their
Hilbert transforms. The solid and dashed lines in the
bottom-right panel correspond to the NC and BP force. For the parameters used
here, the NC and BP force are comparable with the diagonal RN and
friction forces, shown in the two panels on top. The symmetry properties imply
that the NC and the RN term become dominant in the limit of slow
vibrations. The nonequilibrium RN term can be of interest, for example qualitatively 
changing the potential profile\cite{HuMeZeBr.2010,NoPeMa.2011}. Furthermore, 
by combining the RN term with a further contribution, arising from the next order 
in the expansion of the electronic Hamiltonian in powers of the displacements, 
it is possible to construct the full non-equilibrium dynamical response matrix\cite{dundas.2012}. 
Its bias-dependence can compete with the NC force and influence the appearance, 
or otherwise, of waterwheel modes. 

In the present work the RN term only changes quantitatively the results for the 
model used. The RN term will be excluded altogether below, focusing instead on 
the effect of NC and BP force.

\begin{figure}[htpb]
	\begin{center}
		\includegraphics[scale=0.5]{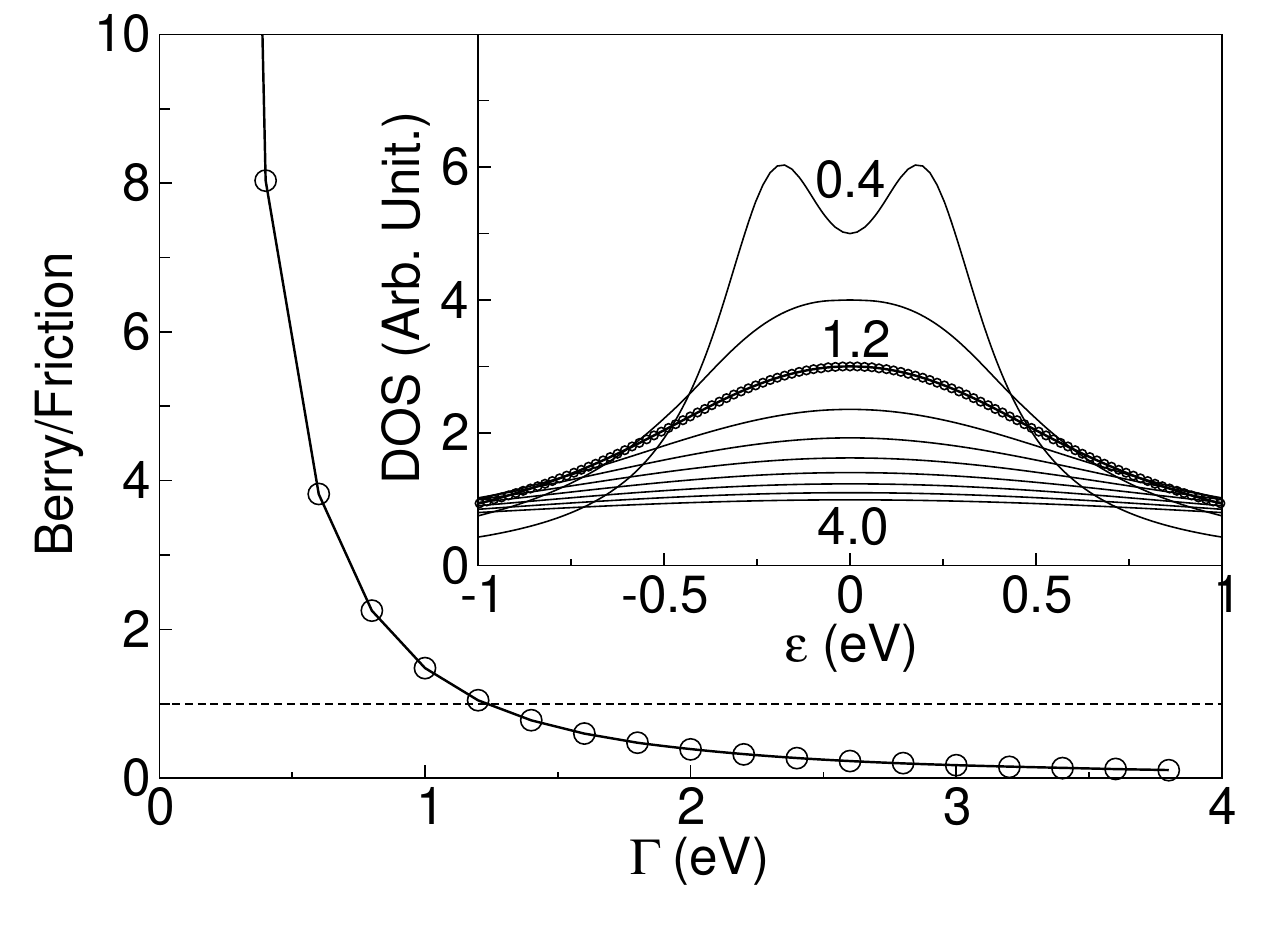}
	\end{center}
	\caption{(main panel) Relative magnitude of the BP force
          compared with the average friction
          $2F_{BP}/({F_{FR}}_{11}+{F_{FR}}_{22})$ as a function of
          level broadening $\Gamma$ with $\hbar\omega_0=0.02$ eV.
          (inset) The electronic DOS function at different
          $\Gamma$.  The BP force equals the friction at $\Gamma=1.2$
          eV.}
	\label{fig:brfr}
\end{figure}

\begin{figure}[htpb]
	\begin{center}
		\includegraphics[scale=0.5]{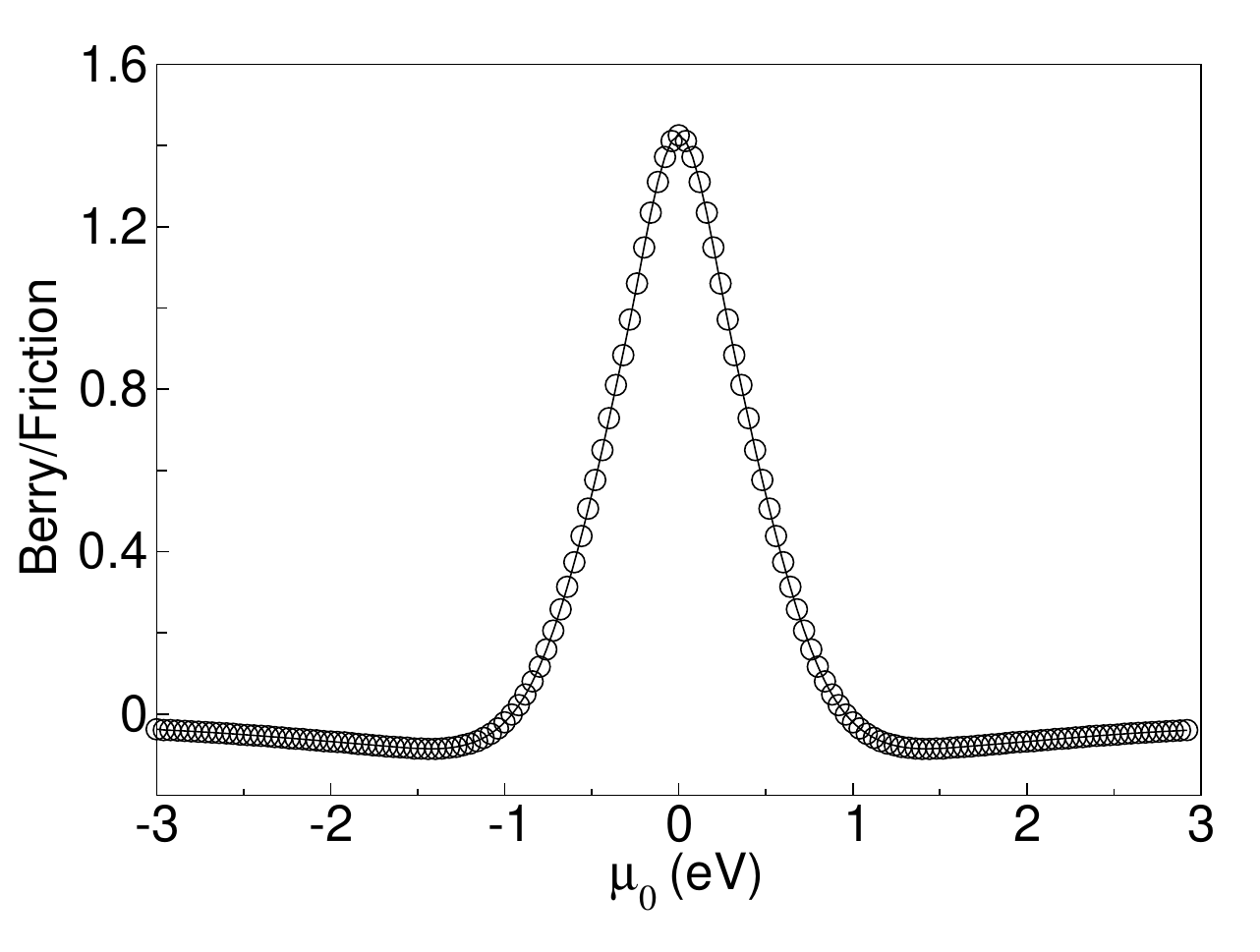}
	\end{center}
	\caption{Relative magnitude of Berry force compared with
	the average friction felt by the two phonon modes ($({F_{FR}}_{11}+{F_{FR}}_{22})/2$) 
	as a function of position of the average electrochemical potential $\mu_0$ at a bias of $1$ V.}
	\label{fig:mudep}
\end{figure}

We already see from the analytical result that the magnitude of the BP force is
directly related to the energy dependence of the electron spectrum.  This is
confirmed numerically in Figs.~\ref{fig:brfr} and \ref{fig:mudep}.  We show in
Fig.~\ref{fig:brfr} the relative magnitude of the BP force compared with the
average diagonal friction for different level broadenings $\Gamma$ at $1$ V with
$\hbar\omega_0=0.02$ eV. The inset shows the left spectral function. We see that
for a range of $\Gamma$, the BP force is of the same
magnitude as the friction. With increasing $\Gamma$ the resonance in the DOS
gets broader, and consequently the BP force gets smaller ($\propto
\partial_\varepsilon R$ as in \Eqref{eq:defco}). In \Figref{fig:mudep} we
vary the energy position of the bias window relative to the peak in the spectral
function. The BP force drops quickly when the bias window moves away from
the peak.

\begin{figure}[htpb]
	\begin{center}
		\includegraphics[scale=0.5]{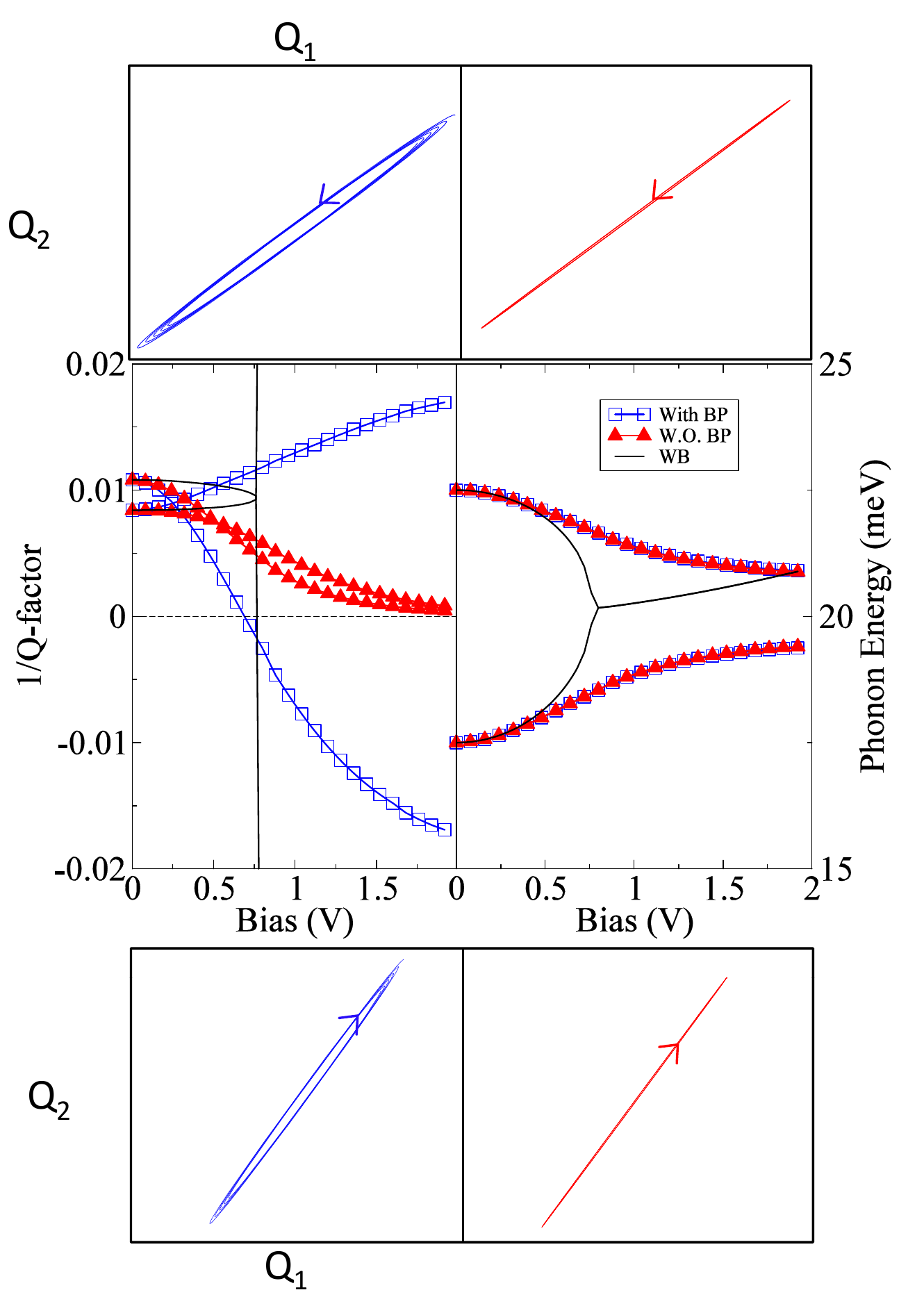}
	\end{center}
	\caption{(Color online) The upper(lower) panels show
	the motion of positive(negative) damped branches calculated at $V=2$ V bias, 
	with and without the BP force in left and right panels, respectively.
The middle panel show the inverse Q-factor to the left, and the phonon energy as a function
	of bias to the right, with and without the BP force.
The solid line is the result obtained from the wideband approximation,
	ignoring the BP force.  Parameters used: $\hbar\omega_0=20$ meV, $\delta=5$ meV.}
	\label{fig:reim1}
\end{figure}

\begin{figure}[htpb]
	\begin{center}
		\includegraphics[scale=0.5]{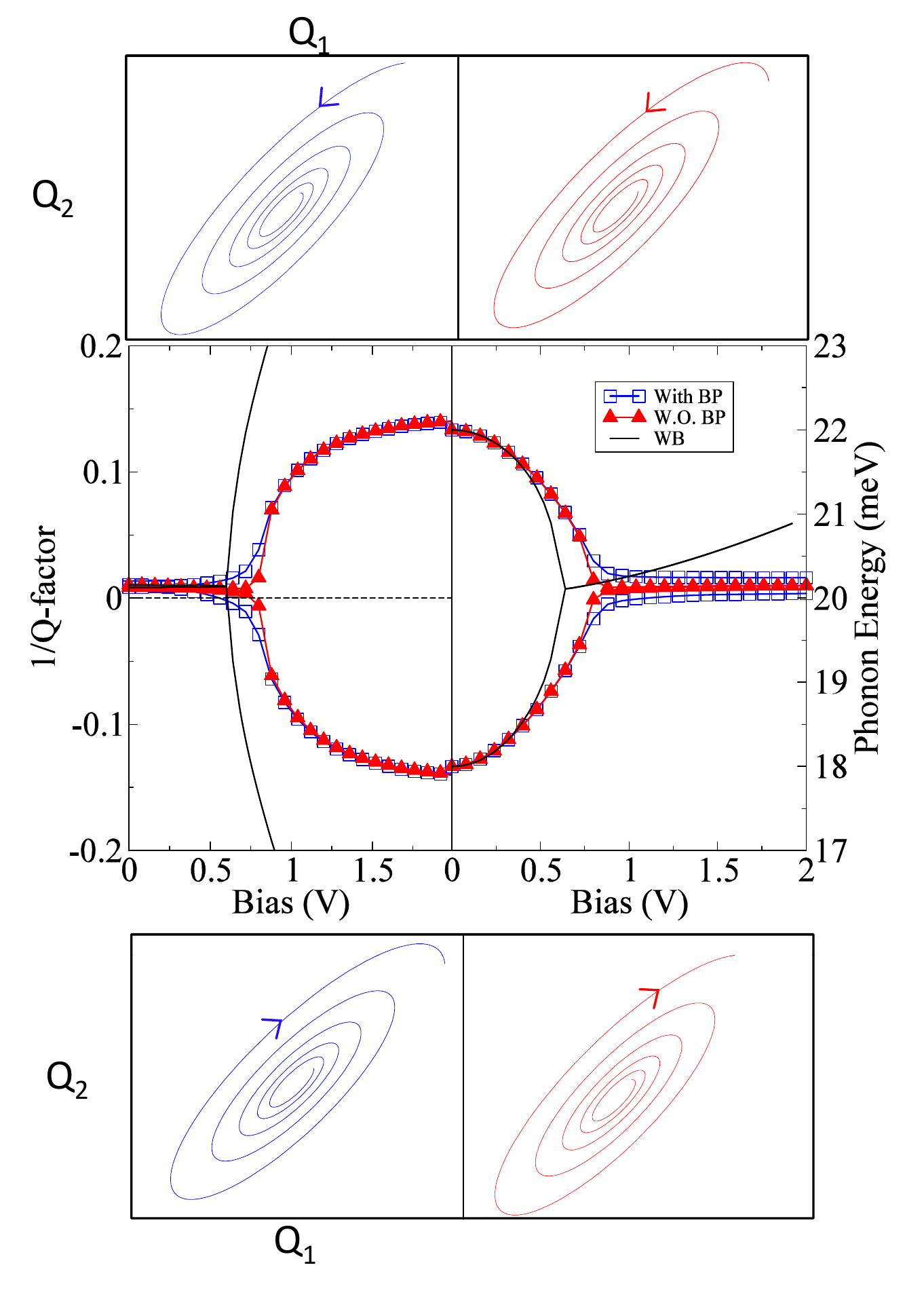}
	\end{center}
	\caption{(Color online) The same as in Fig.~\ref{fig:reim1}, but now with $\delta=4$ meV.}
	\label{fig:reim2}
\end{figure}

Assuming a small detuning of the two harmonic oscillators
$\hbar\omega_\pm=\hbar\omega_0\pm\delta/2$, we now study the bias dependence of their
frequency, and damping described by their inverse Q-factors in Figs.~\ref{fig:reim1} and \ref{fig:reim2}. The
runaway solution is defined at the point where the damping disappears, $1/Q_i = -2{\rm
Im}\,\omega_i/{\rm Re}\,\omega_i=0$. We see that the BP force in general helps the
runaway solution by reducing the threshold bias. This is prominent for larger
detuning. The reason is that it bends the eigenmodes into ellipses so that the
NC force continuously can take energy out of one mode, while pumping energy into the
other. Eventually, this changes the polarization of the harmonic motion from
linear to elliptical (circular) in mode space.

\begin{figure}[htpb]
	\begin{center}
		\includegraphics[scale=0.5]{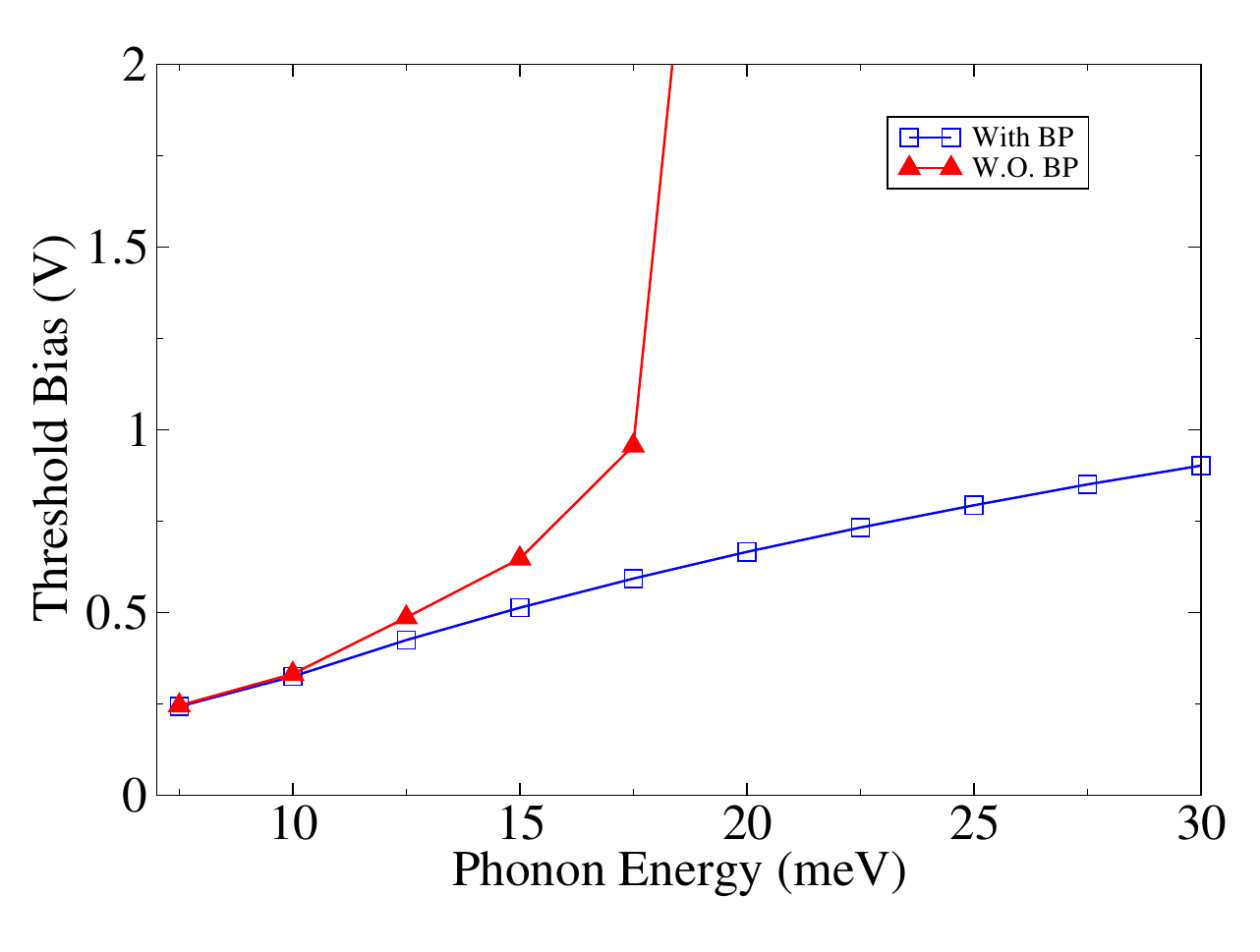}
	\end{center}
	\caption{(Color online) The threshold bias as a function of phonon energy with or without BP force.
The parameters are the same as in Fig.~\ref{fig:reim1}.}
	\label{fig:thres}
\end{figure}

\begin{figure}[htpb]
	\begin{center}
		\includegraphics[scale=0.3]{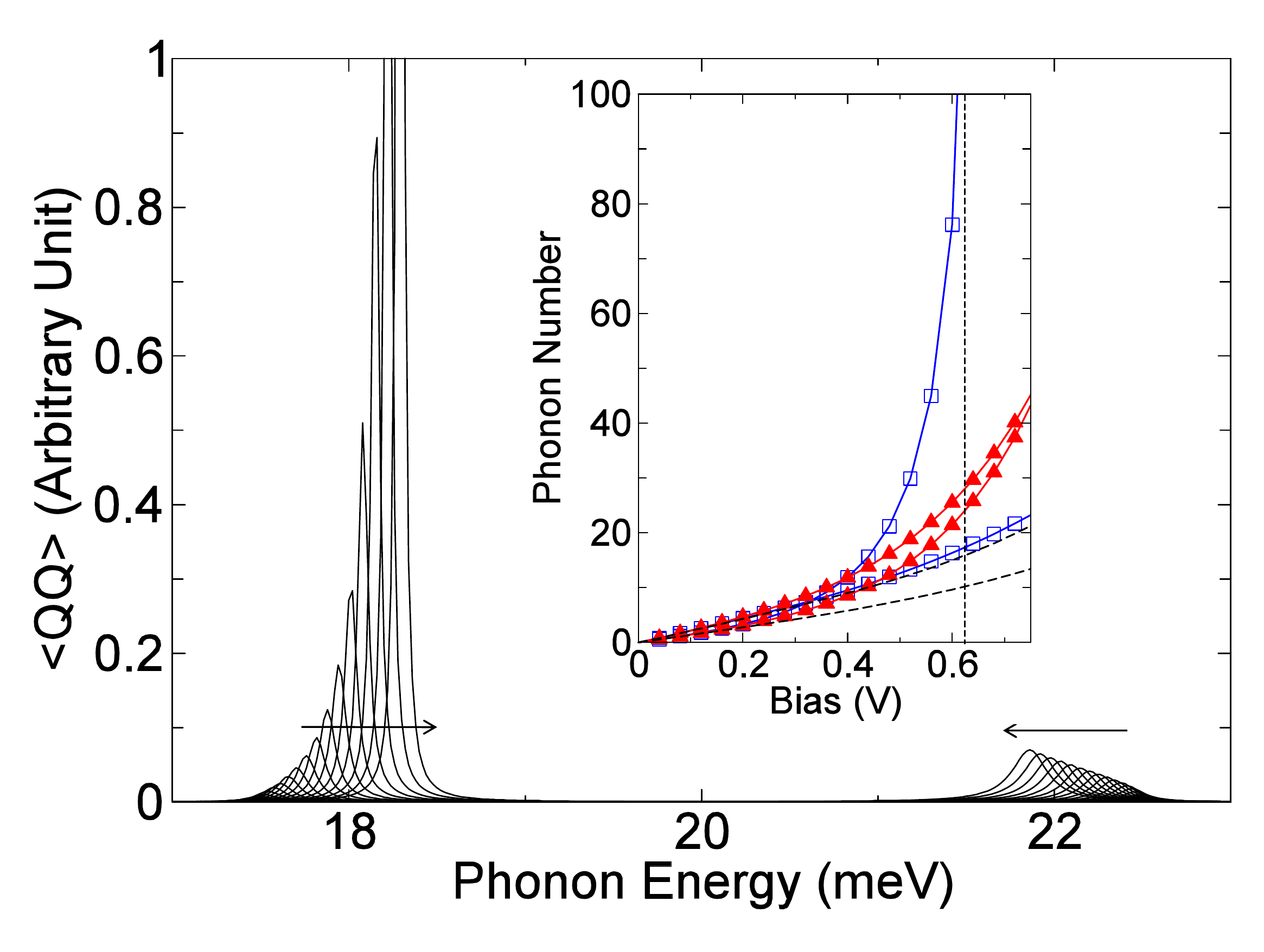}
	\end{center}
	\caption{(Color online) Semi-classical displacement correlation function $\langle
	QQ\rangle$ of the two modes at different bias. The bias increases
	along the arrow shown. (inset) Excited phonon number as a function of bias.
	The dashed curves are the results when only the friction is included. The
	parameters are the same as in Fig.~\ref{fig:reim1}.}
	\label{fig:n}
\end{figure}

Comparing the full calculation with that from
the wideband approximation, we see that they agree well only in the low bias
regime. For large bias, we need to take into account the energy dependence of the
electronic spectral function. The frequency dependence of the threshold bias with
and without the BP force is depicted in Fig.~\ref{fig:thres}.
The divergent behavior of the threshold bias when only the NC force is considered is
due to the finite range of the electron DOS (inset of \Figref{fig:brfr}). Once
the bias is large enough for the bias window to enclose the DOS peaks, 
the NC force will saturate. Further increase of the bias does not
help. But the BP force has an extra linear $\omega$ dependence 
(since it depends on $\dot{Q}$, instead of $Q$),
which becomes important for high frequencies.
The bias dependence of the mode-correlation function and derived 
excited phonon number (cf.~\Eqref{eq:ei}) corresponding to
Figs.~\ref{fig:reim1} is depicted in Fig.~\ref{fig:n}.
Near the threshold bias, the sharp increase of the occupation number of one mode is
a signature of the runaway solution. We will return to the signature in the Raman signal in \Secref{sec:raman}.

\section{Two extensions}
\label{sec:extensions}

\subsection{Adiabatic limit}
\label{sec:adiabatic}

The perturbation approach we have presented, and illustrated with the model
calculation above, is applicable to weak electron-phonon interaction. It is not
restricted to slow ions. Within the same theoretical framework and based on the
Hamiltonian in \Eqref{equa}-\Eqref{equc}, we can carry out an adiabatic expansion,
where the assumption is that the ions are moving slowly, while the
electron-phonon interaction does not have to be
small\cite{BoKuEgVo.2011,BoKuEg.2012}. 

In the limit of small ionic velocities, we expand the displacement in \Eqref{eq:dyson}
at $\tau'$ as follows,
\begin{equation}
	X(\tau') \approx Q({t})\sigma_z + \left(\dot{Q}(t) (t'-t)\sigma_z+\frac{1}{2}\xi(t')I_2\right),
	\label{eq:adia}
\end{equation}
where $\sigma_z$, $I_2$ are the
Pauli and $2\times2$ identity matrix, respectively. Using
Eq.~(\ref{eq:adia}), we can re-group the expansion series in \Eqref{eq:dyson},
and arrive at \begin{eqnarray}
	&&G(\tau,\tau_+) = \mathcal{G}_0(\tau,\tau_+)
	+\sum_k \int_K \mathcal{G}_0(\tau,\tau')M^k \nonumber \\
	 &&\times\left(\dot{Q}_k({t})(t'-t)\sigma_z+\frac{1}{2}\xi(t')I_2\right)G(\tau',\tau_+)d\tau'.
	\label{eq:dyson2}
\end{eqnarray}
Now $\mathcal{G}_0(\tau,\tau_+)\equiv\mathcal{G}_0(\tau,\tau_+;Q(t))$ is the
adiabatic electron Green's function, determined by the instantaneous electronic 
Hamiltonian when the ions are at a given configuration ($Q(t)$).

The force due to the first term in the new Dyson equation now takes the same
form as Eqs.~(\ref{eq:1st}-\ref{eq:fek}), but the non-interacting electron
spectral function $A_\alpha$ is replaced by the adiabatic one,
$\mathcal{A}_\alpha(\varepsilon)\equiv\mathcal{A}_\alpha(\varepsilon;Q(t))$,
which is upto infinite order in $M$,
\begin{equation}
	\mathcal{F}^{(1)}_{k} = - 2\sum_{\alpha} \int\frac{d\varepsilon}{2\pi}\;
	{{\rm Tr}}[\mathcal{A}_\alpha(\varepsilon)M^k]\Delta n^\alpha_F(\varepsilon).
	\label{eq:fek2}
\end{equation}
Its contribution to the forces in the Langevin equation includes {\em both} the
renormalization of the effective potential and the NC force.  To see this, we
assume $Q$ is small, and expand the adiabatic spectral function over $Q$ near
$Q=0$. The first contribution ($Q=0$) is exactly the first order result of the
perturbation calculation (\Eqref{eq:fek0}).  The second contribution (linear in
$Q$) can be split into a symmetric and an anti-symmetric part, which are the
RN and NC force, respectively,
\begin{eqnarray}
  &&\frac{\partial\mathcal{F}^{(1)}}{\partial Q_l} \nonumber\\
  &&=-2\sum_\alpha \int \frac{d\varepsilon}{\pi} {\rm Re}{\rm Tr}\left[ M^k \mathcal{G}_0^r(\varepsilon)M^l\mathcal{A}_\alpha(\varepsilon)\right]\Delta n_F^\alpha(\varepsilon) \nonumber\\
  &&=-2\sum_\alpha \int \frac{d\varepsilon}{\pi} {\rm Re}{\rm Tr}\left[ M^k \mathcal{R}(\varepsilon)M^l\mathcal{A}_\alpha(\varepsilon)\right]\Delta n_F^\alpha(\varepsilon) \nonumber\\
  &&-2\sum_\alpha \int \frac{d\varepsilon}{2\pi} {\rm Im}{\rm Tr}\left[ M^k \mathcal{A}(\varepsilon)M^l\mathcal{A}_\alpha(\varepsilon)\right]\Delta n_F^\alpha(\varepsilon).\nonumber
	\label{eq:feka}
\end{eqnarray}
We can see that in the limit $Q\to0$ they agree with the $\omega\to0$ 
limit of the perturbation results as we should expect. 
Note that above we have treated $M$ as $Q$-independent. 
If we relieve this assumption, then a further term enters 
the RN force\cite{dundas.2012}.

\begin{figure}[htpb]
	\begin{center}
		\includegraphics[scale=0.5]{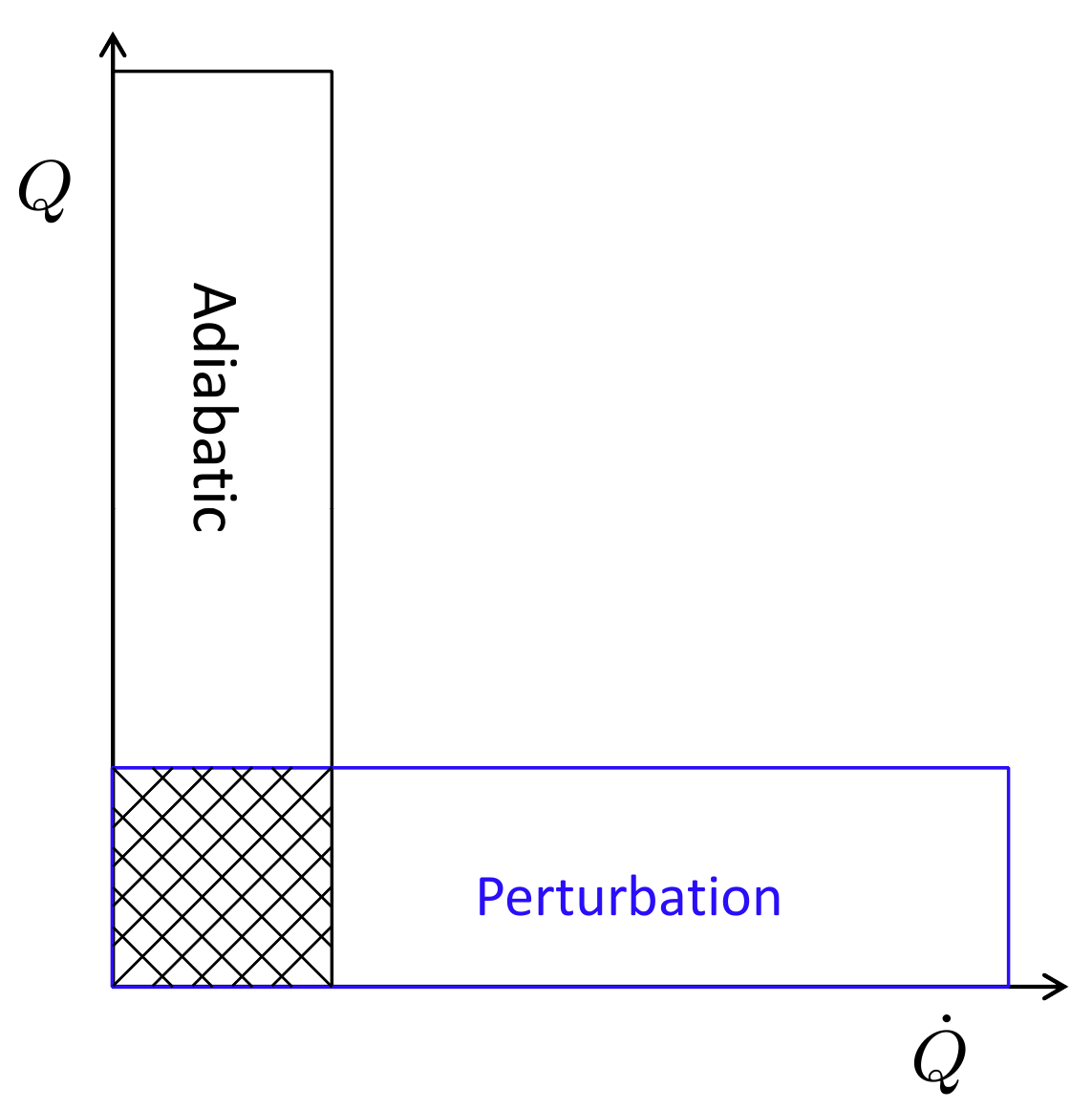}
	\end{center}
	\caption{(Color online) The relation between the perturbation and the adiabatic
	expansion. The perturbation expansion assumes small deviation from the
	equilibrium configuration (small $Q$), while the adiabatic
	approximation assumes slow vibrations (small $\dot{Q}$). In the region
	where both $Q$ and $\dot{Q}$ are small, the two expansions agree with
	each other.}
	\label{fig:adp}
\end{figure}

The effective action from the second term in \Eqref{eq:dyson2} is
given by \Eqref{eq:s2}, with $Q_l(t') \to \dot{Q}_l(t)(t'-t)$,
$A_\alpha(\varepsilon) \to \mathcal{A}_\alpha(\varepsilon)$. Its
contributes to the Langevin equation in terms of the friction, the BP force, and
the noise. To get the expressions for these forces, we need to (1)
take the perturbation results in the $\omega\to0$ limit, (2) replace
$A_\alpha(\varepsilon)$ and $G_0^r(\varepsilon)$ with
$\mathcal{A}_\alpha(\varepsilon)$ and $\mathcal{G}_0^r(\varepsilon)$,
respectively.  Again, the adiabatic results in the $Q\to0$ limit agree
with the perturbation results in the $\omega\to0$ limit.
We conclude this section with the diagram showing the relation between these two
approximations (\Figref{fig:adp}), and noting that the adiabatic approximation 
in principle allows for updating the parameters in the Hamiltonian along the path.

\subsection{Coupling to electrode phonons}
\label{sec:heat}
Actual molecular conductors are coupled also to electrode phonons.  
The energy dissipated by the electrons can be transferred to the 
electrodes via this additional channel\cite{luwang07,PeRoDi.2007,EnBrJa.2009,TsTaKa.2008,ScFrGa.2008}.
The effective action due to linear coupling with a bath of harmonic oscillators is
well-known\cite{Feynman1963,CALE.1983,SC.1982}. 
If we neglect the electron-phonon interaction in the electrodes, 
we can introduce the coupling to electrode-phonons in the Langevin
equation \Eqref{eq:langa} by adding the corresponding phonon self-energy:
$\hat{\Pi}=\hat{\Pi}_e+\hat{\Pi}_{ph}, \Pi^r=\Pi^r_e+\Pi^r_{ph}$. If the
phonon baths are at equilibrium at a given temperature, they have two
effects on the system. One is to modify the effective potential, and the other
is to give rise to dissipation and fluctuating forces, which obey
the fluctuation-dissipation theorem. It is straightforward to include 
a temperature difference between the two phonon baths. A Langevin 
equation including coupling only with phonon baths has been used in 
molecular dynamics simulations to study phonon heat transport\cite{DhRo06,Wa.2007,WaNiJi.2009}. 
It agrees with the Landauer formula in the low temperature limit, 
and with classical molecular dynamics in the high temperature limit.

It is possible to calculate the heat-flux between the central region
and the phonon reservoirs in the electrodes using the self-energies
describing these baths.  If we connect the system to the two
phonon baths ($L$ and $R$), and the nonequilibrium electron bath, 
then the retarded self-energy giving the deterministic force reads, 
$\Pi^r=\Pi^r_e + \Pi^r_L+\Pi^r_R$, and the fluctuating force is,
$f=f_L+f_R+f_e$. When the system reaches a steady state, we can write,
\begin{equation}
	\dot H_{ph} \equiv I_e+I_{L}+I_R\equiv 0,
	\label{eq:bal}
\end{equation}
where $I_\alpha$ is the energy current (power) flowing
\emph{into} the system from each bath, $\alpha$ ($\alpha=L,R,e$).  
Expressions for the power exchange can be found using the forces 
acting between the system and each bath in the Langevin equation,
\begin{equation}
	I_\alpha (t) \equiv -\dot{Q}^T (t) \left(\int^{} \Pi^r_{\alpha}(t-t')Q(t') dt' - f_{\alpha}(t)\right).
	\label{eq:ia}
\end{equation}
Although we employ the harmonic approximation in this paper, this definition
is valid also if there is anharmonic interaction inside the central
region\cite{Wa.2007}. This will be important for example when describing
high-frequency molecular modes which only couple via anharmonic interaction
to the low-frequency phonon modes in the electrodes.
This situation can then be handled by a calculation where the surface
parts of the electrodes are included explicitly in the definition of central region.
We can write the expression for the energy current in frequency domain,
\begin{eqnarray}
	\bar{I}_\alpha &\equiv& \langle I_{\alpha}(t) \rangle \\
&=& -i  \int \frac{d\omega}{2\pi} \omega \left( {\rm tr} \left[ \Pi^r_{\alpha}(\omega) \langle Q{Q}^T\rangle(\omega) - \langle  f_{\alpha} Q^T\rangle(\omega) \right] \right)
\nonumber\,,
	\label{eq:pel0}
\end{eqnarray}
where ${\rm tr}$ denotes trace over phonon degrees of freedom in region $C$.
Using the solution of the Langevin equation, and the noise correlation function,
we can get a compact formula,
\begin{eqnarray}
	\bar{I}_\alpha &=& -i\int \frac{d\omega}{4\pi} \hbar\omega \,{\rm tr}\left[\tilde{\Pi}_{\alpha}(\omega) D^r(\omega)  \hat{\Pi}(\omega)  D^a(\omega)\right.  \nonumber\\
	&&- \left.\hat{\Pi}_\alpha(\omega)D^r(\omega)\tilde{\Pi}(\omega)D^a(\omega)\right].
	\label{eq:fib2}
\end{eqnarray}
This result agrees with NEGF theory\cite{luwang07}, and fullfills the energy
conservation, e.g., $\sum_{\alpha=L,R,e} \bar{I}_\alpha = 0$.  Without the
electron bath and anharmonic couplings, it reduces to the Landauer formula for
phonon heat transport\cite{ReKi98,miles.1999,MiYa03,Mingo06,YaWa.06,WaWaZe.06}.
The formula contains information about the effects of the electrons and the
electronic current on transport of heat to, from and across the central region,
but these effects are beyond the scope of the present paper.  Instead, we now
focus on how the excitation of the localized vibrations by the current affects
their Raman signals.

\section{Raman spectroscopy and correlation functions}
\label{sec:raman}
A central aspect of this work is that we need to find ways to
actually observe the consequences of the current-induced forces.
One promising route\cite{IoShOp.2008,WaCoToNa.2010} is the recent possibility
of doing Raman spectroscopy on single, current-carrying
molecules. In Raman spectroscopy one can deduce the ``effective temperature'' 
(or the degree of excitation) of the various Raman-active vibrational modes of a
system. The semi-classical theory we have introduced is not applicable to Raman spectroscopy, 
since it always gives the same Stokes and anti-Stokes lines, for a reason that will 
be clear in the following. So we are forced to go back to the quantum-mechanical theory.
Mathematically, the Raman spectrum can be written as follows,\cite{SchatzRatnerBook}
\begin{equation}
\mathscr{R}(\omega) = \int a_k \langle x_k(t) x_l(t') \rangle a_l e^{i\omega(t-t')} d(t-t').
\label{eq:raman}
\end{equation}
Here $a_k$ is a vector involving the change in polarizability of the molecule
when its atoms are displaced along the direction $k$ corresponding to the
position operator $x_k$. When coupling with the electrode, $a_k$ could change
due to interaction between the molecule and the electrode\cite{OrGaNi12}. We
will take it as a parameter, and focus on the displacement correlation function
instead.  Now $x_k(t)$ is an operator, and the average in \Eqref{eq:raman} is a
quantum-mechanical one.  Since there is no time ordering in the quantum
correlation function $\langle x_k(t) x_l(t') \rangle$ at the heart of the Raman
expression, it is best implemented in our path integral version with $t'$ in
the upper Keldysh contour and $t$ at the lower contour.  Hence it can be
represented as
\begin{eqnarray}
  \langle x_k(t) x_l(t') \rangle &=& \mathcal{Z}^{-1}\int\mathcal{D} Q \int\mathcal{D} \xi \left(Q_k(t)-\frac{\xi_k(t)}{2}\right )\nonumber\\
  &\times&\left(Q_l(t')+\frac{\xi_l(t')}{2}\right) e^{\frac{i}{\hbar} S_{eff}(Q,\xi)},
\end{eqnarray}
where the effective action can be found in formula (\ref{eq:s2}),
\begin{eqnarray}
  S_{eff}(Q,\xi) &=& -\frac{1}{2}\int \frac{d\omega}{2\pi}\left[ Q^\dagger(\omega) L^\dagger (\omega) \xi(\omega) + \xi^\dagger(\omega) L(\omega) Q(\omega)\right. \nonumber \\
  && \left. -i \xi^\dagger(\omega) \hat{\Pi}(\omega) \xi(\omega) \right],
\end{eqnarray}
and $\mathcal{Z}$ is a normalization factor. 
The Raman spectrum thus has four contributions. A classical
contribution, $\mathscr{R}_{QQ}(\omega)$, proportional to the average of
$Q_k(\omega) Q_l(\omega)^*$, two quantum corrections,
$\mathscr{R}_{Q\xi}(\omega)$ and $\mathscr{R}_{\xi Q}(\omega)$, 
proportional to the averages of $Q_k(\omega) \xi_l(\omega)^*$ and 
$\xi_k(\omega)Q_l(\omega)^*$, and finally a contribution $\mathscr{R}_{\xi\xi}(\omega)$
proportional to $\xi_k(\omega) \xi_l(\omega)^*$. The calculation of
these averages involves simple Gaussian integrals, and the results are
\begin{eqnarray}
\mathscr{R}_{QQ}(\omega) &=& a_k [\hbar D^r(\omega) \hat{\Pi}(\omega) D^a(\omega)]_{kl} a_l \\
\mathscr{R}_{Q\xi}(\omega) &=& \frac{i}{2} a_k \hbar D^r_{kl}(\omega) a_l  \\
\mathscr{R}_{\xi Q}(\omega) &=&  -\frac{i}{2} a_k \hbar D^a_{kl}(\omega) a_l  \\
\mathscr{R}_{\xi\xi}(\omega) &=& 0.
\end{eqnarray}

These functions are dominated by the properties close to the poles of
$L^{-1}$. Let us first consider the case of one mode in thermal
equilibrium. In this case $L(\omega)$ is a simple function which can
be approximated as
\begin{equation}
L(\omega) = -\omega^2 + \omega_0^2 - i\eta\omega.
\end{equation}
In the same approximation $\hat{\Pi}(\omega)$ is controlled by the
fluctuation-dissipation theorem and becomes ($\beta = \hbar/(k_B T)$),
\begin{equation}
\hat{\Pi}(\omega) = \eta\omega\coth\left(\frac{\omega\beta}{2}\right).
\end{equation}
The classical contribution is now
\begin{equation}
\mathscr{R}_{QQ}(\omega) = a^2 \frac{\eta\hbar\omega\coth\left(\frac{\omega\beta}{2}\right)}{(-\omega^2+\omega_0^2)^2+\eta^2\omega^2}.
\end{equation}
This function yields a Raman signal which is symmetric in $\omega$.
It has Lorentzian peaks at $\omega=\pm \omega_0$ of width 
$\eta/(2\omega_0)$, and with strengths given by its area 
(integral over $\omega$), $a^2 \hbar\pi/(2\omega_0)\coth(\beta\omega_0/2)$. 
Note that the strength is proportional to temperature in the high-temperature
limit. It is also important to note that the strength does not depend
on the damping, $\eta$. The self-energy, $\hat{\Pi}(\omega)$, contributes a factor
$\eta$ to the strength, but the $\omega$-integration contributes a factor
$\eta^{-1}$, hence canceling the $\eta$ dependence. The physics of
this is the fluctuation-dissipation theorem of equilibrium: a
smaller damping should give higher oscillation amplitudes were it not
for the associated decrease in fluctuations in the environment.

The quantum correction $\mathscr{R}_{Q\xi}(\omega)+\mathscr{R}_{\xi Q}(\omega)$
is proportional to the imaginary part of $L^{-1}$. In the above
approximation it becomes,
\begin{equation}
\mathscr{R}_{Q\xi}(\omega) + \mathscr{R}_{\xi Q}(\omega) 
= a^2 \frac{\eta\hbar\omega}{(-\omega^2+\omega_0^2)^2+\eta^2\omega^2}.
\end{equation}
This contribution breaks the $\omega\rightarrow -\omega$ symmetry, and
is hence responsible for the different strengths of the Stokes(phonon emission) and
anti-Stokes(phonon absorption) lines. This term also exhibits peaks at $\pm\omega_0$, 
with strengths $\pm a^2\hbar\pi/(2\omega_0)$.

The ratio, $r$, of strengths of the anti-Stokes and Stokes lines becomes,
\begin{equation}
\label{eq:SAratio}
r = \frac{\coth\left(\frac{\beta\omega_0}{2}\right)-1}{\coth\left(\frac{\beta\omega_0}{2}\right)+1} = e^{-\beta\omega_0}.
\end{equation}
In equilibrium this ratio is used to measure the temperature employing 
the particular mode excitation, but has also been used to estimate 
an effective temperature of the different modes out of equilibrium\cite{WaGrLe.2007,IoShOp.2008}.

Next we discuss the situation in the presence of electronic current and the 
derived forces. We consider a situation with two modes close in frequency, 
which are coupled by the current. Now, the function $L(\omega)$ is a $2\times 2$ matrix. If the
NC and BP forces are represented by constants, ${n}$ and ${b}$, respectively,
then $L(\omega)$ becomes
\begin{equation}\label{TwoCoupledModes}
L(\omega) = \left(\begin{array}{cc} -\omega^2 + \omega_1^2 - i\omega \eta & n - i \omega {b}  \\
 -{n}  +i \omega {b} & -\omega^2 + \omega_2^2 - i\omega \eta\end{array} \right ).
\end{equation}
Here we have made the simplifying assumption that the friction, $\eta$,
is the same for the two modes. The two mode
frequencies $\omega_1$ and $\omega_2$ have a
difference $\delta = \omega_2-\omega_1$.
Comparing with the expressions Eqs.~(\ref{eq:defco0}), (\ref{eq:defco}), 
the parameters ${n}$ and ${b}$ are both linearly dependent on the applied voltage $V$.
If this voltage is sufficiently large, there is a possibility that one of the
eigenmodes of $L(\omega)$ will have a vanishing imaginary part at a
critical voltage $V_c$, and we obtain a runaway mode.

In the following we present plots of the resulting Raman spectra as a
function of voltage approaching $V_c$ from below. The function
$\hat{\Pi}(\omega)$ describing the fluctuating forces is both
temperature- and voltage-dependent. However, nothing dramatic happens
at the critical voltage, so it will be taken to be a temperature and
voltage independent matrix with values of the order of
$\eta\omega$. As in the equilibrium case, an increasing lifetime
of the mode will lead to stronger oscillation amplitudes, but out of
equilibrium this need {\em not} be counteracted by decreasing fluctuations 
in the environment. The result is a strong increase in the strength of the
runaway mode, both for the Stokes and the anti-Stokes line.

\begin{figure}[htpb]
	\begin{center}
		\includegraphics[scale=0.4]{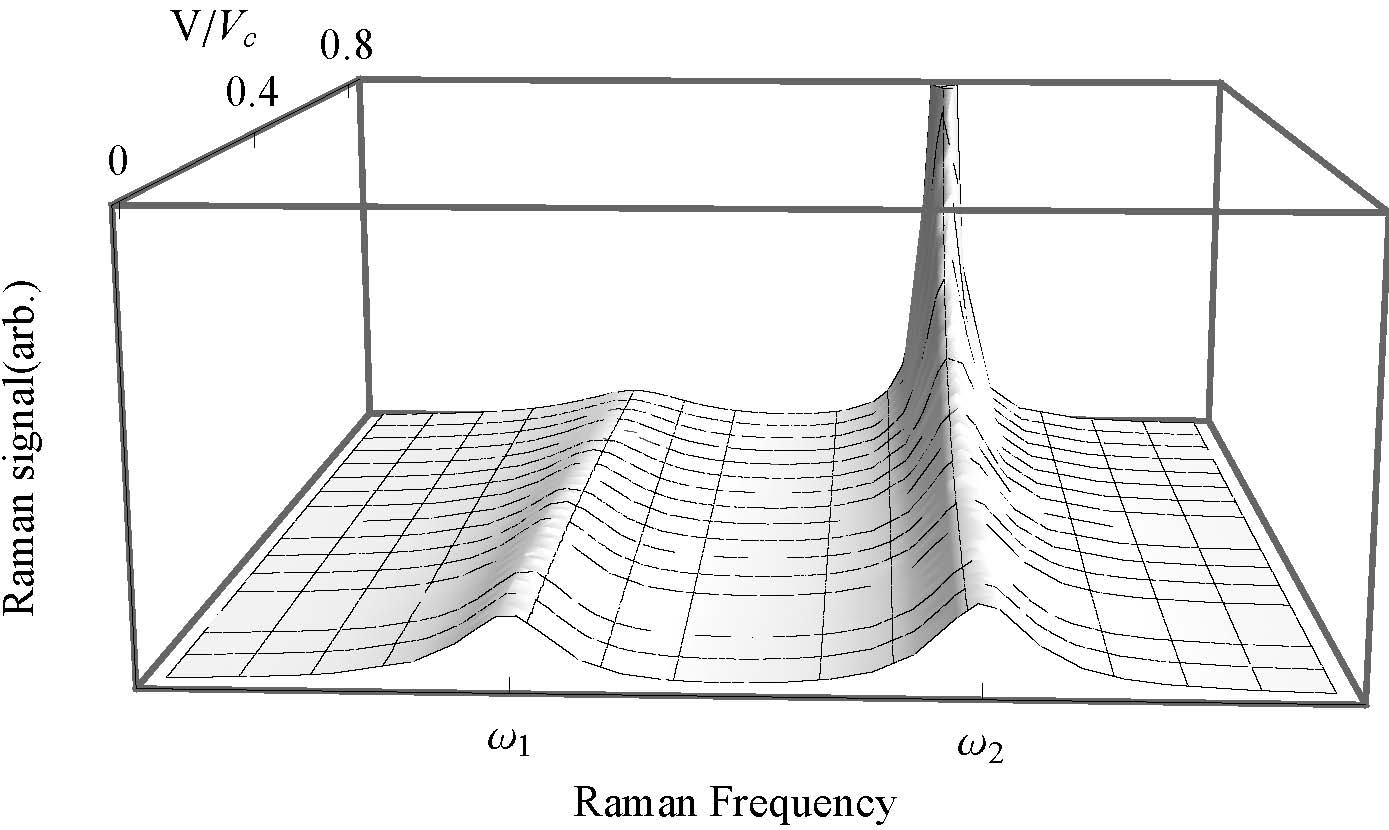}
	\end{center}
	\caption{(Color online) The Stokes lines of two coupled modes for various values of the applied voltage in units of the critical voltage.
Parameter values are $\omega_1 = \bar{\omega}-\delta/2$, $\omega_2=\bar{\omega}+\delta/2$
with $\delta = 0.2 \bar{\omega}$, $\eta = 0.05 \bar{\omega}$,
${b}/{n} = 0.6/\bar{\omega}$.}
	\label{fig:rs1}
\end{figure}

\begin{figure}[htpb]
	\begin{center}
		\includegraphics[scale=0.4]{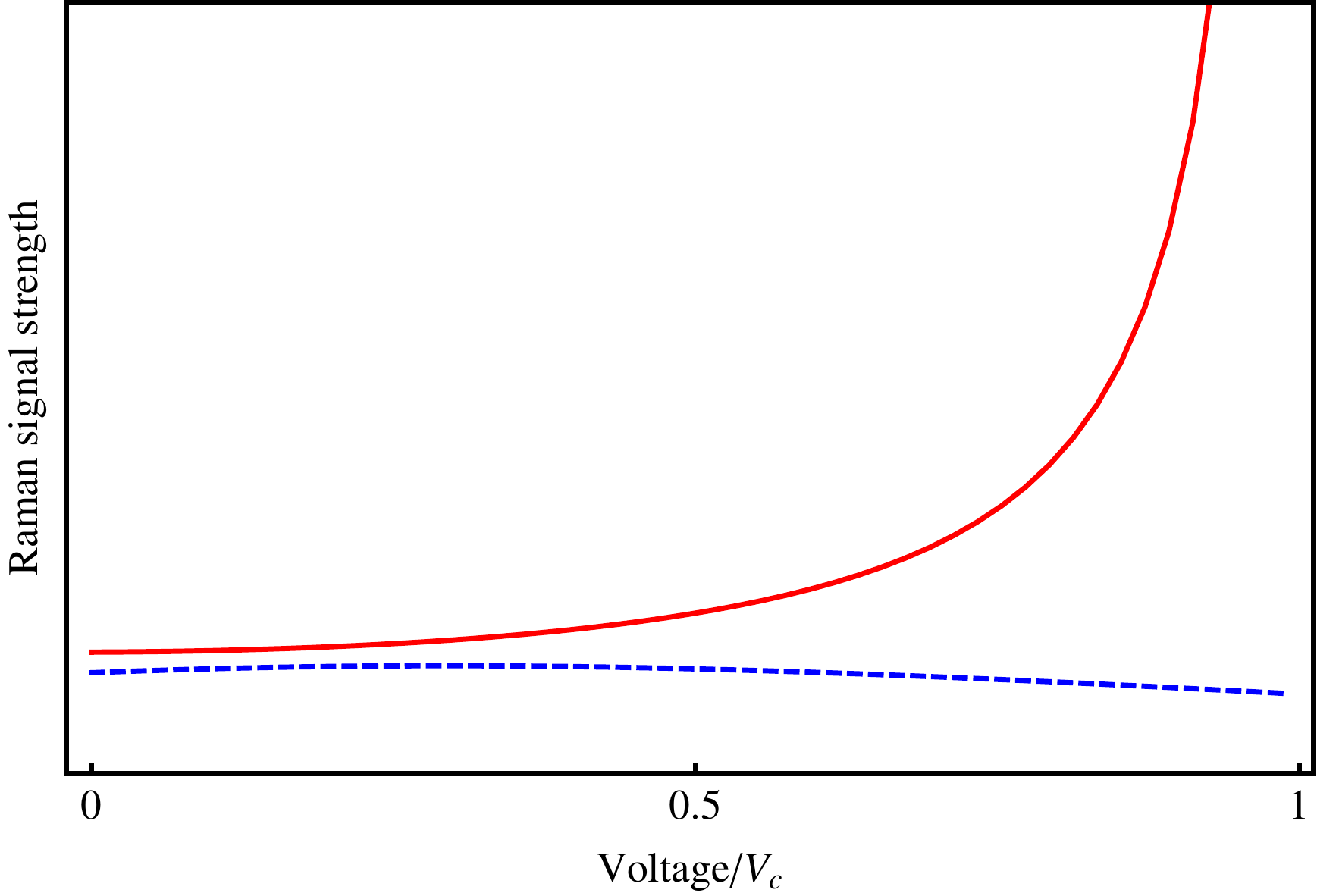}
	\end{center}
	\caption{(Color online) Strength of the two Stokes lines as a function of voltage.
 Solid line corresponds to the ``runaway'' mode at $\omega_2$, and dashed line to the ``cooling'' mode at $\omega_1$.}
	\label{fig:rs2}
\end{figure}

\begin{figure}[htpb]
	\begin{center}
		\includegraphics[scale=0.4]{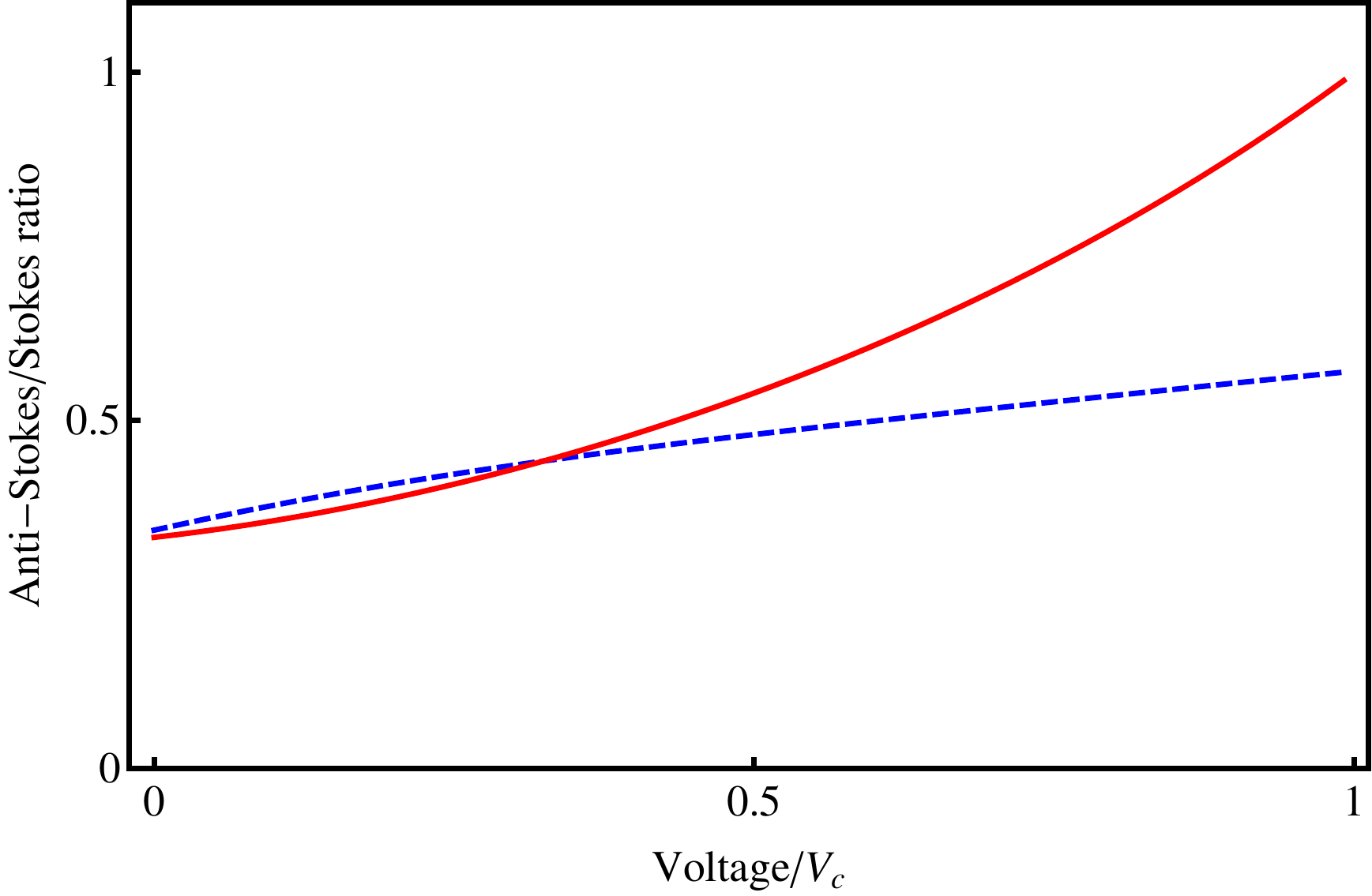}
	\end{center}
	\caption{(Color online) Ratio of strengths of the anti-Stokes and Stokes lines as a function of voltage.
Parameters are as in \Figref{fig:rs1}. Solid line corresponds to the ``runaway'' mode at $\omega_2$, and dashed line to the ``cooling'' mode at $\omega_1$.
}
	\label{fig:rs3}
\end{figure}

Figure \ref{fig:rs1} shows the Stokes lines for two modes close in
frequency and coupled by the NC and BP forces, for the system described by
Eq.~(\ref{TwoCoupledModes}). In the following graphs the parameters are,
$\omega_1 = \bar{\omega}-\delta/2$, $\omega_2=\bar{\omega}+\delta/2$ with
$\delta = 0.2 \bar{\omega}$, $\eta = 0.05 \bar{\omega}$, ${b}/{n} = 0.6/\bar{\omega}$, for
which the critical value of the strength of the NC force will be ${n}_c = 0.124
\bar{\omega}^2$, when the voltage reaches its critical the value, $V_c$.  We see
that the mode at $\omega_2$ is increasing in strength while its frequency is
shifting slightly down.  By contrast the frequency of the mode at $\omega_1$ is
moving upwards while the strength is decreasing.  If we compare to the standard
one-mode, equilibrium theory of Raman lines, we would say that one mode heats
up, while the other cools down. This is seen clearly in \Figref{fig:rs2}, where
the strength of the two modes is plotted as a function of voltage. Here we see
a small cooling of one mode, while the other mode has a diverging temperature,
as the voltage approaches $V_c$. Alternatively, one could use the ratio
of the anti-Stokes/Stokes strengths in combination with \Eqref{eq:SAratio} to
determine an effective mode-temperature, as shown in \Figref{fig:rs3}.
Interestingly, we cannot see the cooling effect from the anti-Stokes/Stokes ratio.

This could be qualitatively understood as follows. In principle, Eq. (88) only holds
at equilibrium, which is a result of the fluctuation-dissipation theorem.
Under nonequilibrium conditions, discrepancies between different ways of defining 
an effective temperature may be expected.
Specifically, in the case studied here, the bias modifies the rates for phonon 
emission and absorption. This can be inferred from the change of
peak broadening in the phonon correlation function, as well as from the shift 
of the peaks, with bias. Part of this bias-dependent
effect is lost, when we form the anti-Stokes/Stokes ratio.
But it is included when we look at the bias-dependent strength of the Stokes lines.


\section{Concluding remarks}
\label{sec:conc}
In this paper we have derived a semi-classical Langevin equation describing the motion
of the ions, in the harmonic approximation, in nanoscale conductors including both the 
effective action from the current-carrying electrons and the coupling to the phonon baths
in the electrodes. Joule heating and current-induced forces are described on an equal 
footing by this methodology. We derive a convergent expression for the BP force, 
removing the infrared divergence in our previous result\cite{JMP.2010}. The importance 
of the BP force in relation to the stability of the device is further highlighted in a
two-level model system. Using the same model we show the signature of the current-induced 
runaway mode excitation in the Raman spectroscopy of a current-carrying molecular
conductor\cite{IoShOp.2008,WaCoToNa.2010}.

We should mention that the harmonic approximation used here breaks down when
the phonons are highly excited. In that case the anharmonic coupling between 
different phonon modes becomes important. Analysis in this regime relies e.g. 
on molecular dynamics simulation. To take the nonlinear potential approximately 
into account in the present method, we only need
to replace the force from the harmonic potential $-KQ(t)$ with the full
potential $-\partial_Q V(Q)$. Thus the full Langevin equation has the following
advantages: first, we can include the highly nonlinear ion potential, which is
necessary to simulate bond-breaking processes; second, a crucial part of the
quantum-mechanical motion is included in the dynamics. For example, we recover
the quantum-mechanical results for Joule heating and for heat transport in 
the harmonic limit. This enables the study of phonon heat
transport\cite{WaWaLu07}, or thermoelectric\cite{DuDi11} transport using the
same Langevin equation including the electron-phonon interaction, and
is an interesting topic for future research.

The backaction of the runaway vibrations on the electrons is another possible
extension of the present study. There are at least two effects to address. The
first one is the adiabatic change of the electronic structure. The purpose of
the adiabatic extension in Sec.~\ref{sec:adiabatic}  is to include this effect.
The second effect is the inelastic electrical current, which becomes important
when the electron mean free path is comparable to the device length.

A number of problems need to be faced when implementing the approach within the
framework of density functional theory.  First, the ions may be driven away from 
their equilibrium positions by the current. The electronic structure and 
electron-phonon coupling depend on the ionic positions, and one may need to 
update the electronic friction and noise correlation function throughout the 
molecular dynamics simulation. This introduces a technical problem, in addition to 
the computational challenge, which is how to generate colored noise when its 
correlation function is time-dependent. Second, calculation of the convolution 
kernel in the Langevin equation is time-consuming considering the many time-steps 
needed to sample the dynamics. For the electron bath in the wideband limit the
convolution transforms to time-local forces. But the time-scale of the phonon
bath is typically comparable to that of the system, and we cannot use the wideband
approximation. One possible solution could be to include more ions from the baths 
into the dynamical region and approximate the coupling to the external
phonon bath with time-local forces. Intuitively, with more ions included, the
approximate central system will be closer to the actual system under study. 
Work to overcome these difficulties is underway.

Electron-nuclear dynamics is a broad problem, relevant to many fields.
A central challenge is how to take account of electron-nuclear correlation. 
The simplest form of nonadiabatic dynamics - the Ehrenfest approximation - fails
precisely there: it does not take into account spontaneous phonon emission
by excited electrons and consequently the Joule heating effect. The appeal 
of Ehrenfest dynamics is its conceptual simplicity derived from the classical 
treatment of the nuclei. The Langevin approach retains this key element, while 
rigorously reinstating the vital missing ingredient: the fluctuating  
forces - with the correct noise spectrum - exerted by the electron gas and 
responsible for the return of energy from excited electrons to thermal vibrations. 
Thus, in addition to its capabilities as a method for nonadiabatic dynamics, 
this approach can be very helpful conceptually, by explicitly quantifying effects 
that are intuitive but whose physical content can sometimes remain hidden from view.

Although our discussion here is in the context of molecular electronics, 
we believe that the predictions based on the simple model can be important
for other interesting physical systems, for instance nano-electromechanical oscillators
(NEMS) coupling with an atomic point contact\cite{kemiktarak_radio-frequency_2007,poggio_off-board_2008}.  
When the current through an atomic point contact is used to measure the motion of an oscillator, 
the measurement imposes quantum backaction on the oscillator.
A similar Hamiltonian describes this quantum backaction quite
well\cite{HuMeZeBr.2010,bennett_scattering_2010,NoPeMa.2011}. If we now
couple two identical oscillators, or two nearly-degenerate eigenmodes by the
point contact, it seems to be experimentally feasible to detect the polarized motion
predicted here\cite{perisanu_beyond_2010,PhysRevLett.106.094102}.

A neighbouring field, where much larger size- and time-scales come into 
play, and where Langevin dynamics is an important line of approach,
is the simulation of radiation damage\cite{chris.2010}. It is hoped that 
the present discussion will be of interest to that community.

\section*{Acknowledgements}
We thank Professor Jian-Sheng Wang for discussions.
Financial support by the Lundbeck foundation (R49-A5454) is gratefully acknowledged.
D.D. and T.N.T. are grateful for support from the Engineering and Physical Sciences 
Research Council (grant EP/I00713X/1). We thank Andrew Horsfield and Jorge Kohanoff
for many discussions.

\appendix
\section{Connection with Nano Lett., {\bf 10}, 1657 (2010) }
\label{sec:nl}
In Ref.~\onlinecite{JMP.2010}, we carried out an adiabatic expansion in a slightly
different way from that in Sec.~\ref{sec:adiabatic}. We obtained an infrared 
divergence in the expression for the BP force, and used the largest phonon frequency as a
cutoff. Here we show that by introducing a small lifetime broadening to the scattering
eigenstate ($\gamma$), we can remove the divergence, and get the same result as shown in
this paper. We start from the expression for the Berry force in
Ref.~\onlinecite{JMP.2010}, and write it as
\begin{eqnarray}
	\mathcal{B}_{}=-\lim_{\gamma\to0}\int d\omega\frac{{\rm Im}\Delta\Lambda_{}(\omega)}{(\omega-i\gamma)^2}.
\end{eqnarray}
We first do a partial integration to get
\begin{eqnarray}
	\mathcal{B}_{}=-\lim_{\gamma\to0}\int \frac{d\omega}{\omega-i\gamma} \partial_\omega{\rm Im}\Delta\Lambda_{}(\omega).
\end{eqnarray}
From the $\omega$-dependent part of ${\rm Im}\Delta\Lambda(\omega)$(\Eqref{eq:dl})
\begin{eqnarray}
	&&\lim_{\gamma\to0}{\rm }\int d\omega \frac{\partial_\omega \mathcal{A}(\varepsilon_+)+\partial_\omega \mathcal{A}(\varepsilon_-)}{\omega-i\gamma}
	\sim-4\pi\hbar\partial_\varepsilon {\cal R}(\varepsilon),\nonumber\\
\end{eqnarray}
which gives
\begin{eqnarray}
	\mathcal{B}_{} &\approx& 2\hbar\sum_{\alpha}\int_{}^{}\frac{d\varepsilon}{\pi}\; \Delta n_F^\alpha(\varepsilon)\nonumber \\
&\times&{ {\rm Im} {\rm Tr}}[M\mathcal{A}_\alpha(\varepsilon)M\partial_\varepsilon{\rm Re}{\cal R}(\varepsilon)].
\end{eqnarray}
This agrees with the result in Sec.~\ref{sec:adiabatic}.

In the wideband limit, ignoring the $\omega$ dependence of ${\rm Im}\Delta\Lambda$, we get
\begin{eqnarray}
	\mathcal{B} \approx \frac{2\hbar eV}{\pi\Omega_c}\chi^-,
	\label{eq:dbp2}
\end{eqnarray}
where $\Omega_c$ is an upper bound on the electron-hole pair excitation. 
In Ref.~\onlinecite{JMP.2010}, we used the largest phonon frequency instead, which over-estimates 
the effect of BP force. If as a conservative estimate of $\Omega_c$ we take a typical hopping matrix element,
then using $\dot{Q}/Q \sim \omega$ we get the estimate in \eqref{eq:bpnc}.

\section{Alternative derivation of the Langevin equation}
\label{app:qccor}
In this Appendix, we give an alternative way of arriving at the generalized 
Langevin equation (\Eqref{eq:langa}). The derivation here is meant to be intuitive
rather than theoretically rigorous. Our starting point is the equation of
motion for the (mass-normalised) displacement operator $u$,
\begin{equation}
	\ddot{x} = -Kx +F_e,
	\label{eq:eom1}
\end{equation}
where we define the electronic force operator
\begin{equation}
	F_e = -\frac{\partial H_e(x)}{\partial x}.
	\label{eq:eom2}
\end{equation}
With the help of Green's functions, Eq.~(\ref{eq:eom1}) can be cast into the following form, for each degree of freedom $k$,
\begin{eqnarray}
	\ddot x_k &=& -\sum_jK_{kj}x_j+i\hbar{\rm Tr}\left[M^kG^<(t,t_+)\right]+{f}_k(t).
	\label{eq:eomwe}
\end{eqnarray}
We have defined the noise operator
\begin{equation}
	{f}_k(t) = i\sum_{m,n}M_{mn}^k\left(ic_m^\dagger(t)c_n(t)-\hbar G^<_{nm}(t,t_+)\right),
	\label{eq:xi}
\end{equation}
and the lesser Green's function is $G_{nm}^<(t,t_+) = (i/\hbar)\langle
c^\dagger_m(t_+)c_n(t)\rangle$. The quantum average $\langle \dots \rangle$
is over the electronic environment, which need not be in equilibrium.

From \Eqref{eq:dyson} to second order in $M$,
\begin{eqnarray}
	&&G^<(t,t_+) = G^<_0(t,t_+)\\
	&+&\sum_k\int G_0(t,t')M^kx_k(t')G_0^<(t',t_+)dt'\nonumber\\
	&-&\sum_k\int G_0^<(t,t')M^kx_k(t')\bar{G}_0(t',t_+)dt'.\nonumber
	\label{eq:12nd}
\end{eqnarray}
Using this in \Eqref{eq:eomwe}, we get
\begin{eqnarray}
	\ddot x_k &=& -\sum_jK_{kj}x_j+i\hbar{\textrm{Tr}}[M^kG^<_0(t,t_+)] \nonumber\\
	&&- \sum_j\int \Pi^r_{kj}(t,t')x_j(t')dt'+{f}_k(t),
	\label{eq:eomwe2}
\end{eqnarray}
which is of the same form as \Eqref{eq:langa}.

Now consider the time correlation of the noise operator $f$,
\begin{eqnarray}
	\langle f_i(t)f_j(t')\rangle 	&=& \hbar^2{\textrm{Tr}}[M^iG_0^>(t,t')M^jG_0^<(t',t)]\nonumber\\
	&=& i\hbar\Pi^>_{ij}(t,t')\,,
	\label{eq:cor1}
\end{eqnarray}
and
\begin{eqnarray}
	\langle f_j(t')f_i(t)\rangle = i\hbar\Pi^<_{ij}(t,t')\,.
	\label{eq:cor2}
\end{eqnarray}
As expected, the quantum-mechanical noise operators at different times does not commute.

To go to the semi-classical approximation, we take the
classical noise correlation as the average of Eqs.~(\ref{eq:cor1}) and
(\ref{eq:cor2}),
\begin{eqnarray}
	&&\langle f_i(t)f_j(t')\rangle_{c} =\langle f_j(t')f_i(t)\rangle_{c}\\
	&=& \frac{1}{2}(\langle f_i(t)f_j(t')\rangle+\langle f_j(t')f_i(t)\rangle),\nonumber
	\label{eq:corc}
\end{eqnarray}
where $\langle\dots\rangle_c$ denotes a classical statistical average. Now
the noise spectrum becomes ``classical'', in the sense that its time
correlation function is real. If the potential is anharmonic, the
right side of \Eqref{eq:eomwe2} will contain terms of higher order in
$u$. The equations of motion of these higher-order terms form an infinite 
hierarchy. Then the semi-classical approximation will have to 
involve a truncation procedure. However, this is a separate problem,
beyond the scope of this paper.

We stress again that after making the semi-classical approximation,
it is still possible to calculate the quantum-mechanical average of
two displacement operators at equal times $\langle u_i(t)u_j(t)\rangle$ 
within the harmonic approximation, from the semi-classical Langevin equation,
leading to the correct quantum-mechanical vibrational 
energy and steady-state transport properties.

\section{The phonon self-energy}
\label{sec:se}
The phonon self-energies correspond to the bubble diagram in \Figref{fig:bubble}:
\begin{eqnarray}
	\Pi^{<,>}_{kl}(t-t') &=& -i\hbar\, {\rm Tr}\left[M^kG^{<,>}_{0}(t-t')M^lG^{>,<}_{0}(t'-t)\right],\nonumber\\
	\Pi^{r,a}_{kl}(t-t') &=& -i\hbar\, {\rm Tr}\left[M^kG^{r,a}_{0}(t-t')M^lG^{<}_{0}(t'-t)\right]\nonumber\\
	&&-i\hbar\,{\rm Tr}\left[M^kG^<_{0}(t-t')M^lG^{a,r}_{0}(t'-t)\right].
	\label{eq:ser}
\end{eqnarray}


\end{document}